\documentclass[preprint,showpacs,preprintnumbers,amsmath,amssymb,superscriptaddress,nofootinbib]{revtex4}
\usepackage{graphicx,color}
\usepackage{amsmath,amssymb}
\usepackage{url}
\usepackage{epstopdf,slashed}

\newcommand{\be}{\begin{equation}}
\newcommand{\ee}{\end{equation}}
\newcommand{\bea}{\begin{eqnarray}}
\newcommand{\eea}{\end{eqnarray}}

\newcommand{\crn}{\nonumber \\}

\newcommand{\fr}{\frac}

\newcommand{\bc}{\begin{center}}
\newcommand{\ec}{\end{center}}

\newcommand {\ba}{\begin{array}}
\newcommand {\ea}{\end{array}}
\newcommand{\ben}{\begin{enumerate}}
\newcommand{\een}{\end{enumerate}}

\usepackage{bm}
\usepackage{dcolumn}

\begin{document}

\title{ $(g-2)_{e,\mu}$ and Lepton flavor violating  decays  in a left-right model }

\author{L.T. Hue \footnote{corresponding author}} \email{lethohue@vlu.edu.vn} 
\affiliation{Subatomic Physics Research Group, Science and Technology Advanced Institute, Van Lang University, Ho Chi Minh City, Vietnam}
\affiliation{Faculty of Applied Technology, School of Technology,  Van Lang University, Ho Chi Minh City, Vietnam}
\author{Khiem Hong Phan}\email{phanhongkhiem@duytan.edu.vn}
\affiliation{\it Institute of Fundamental and Applied Sciences, Duy Tan University,
	Ho Chi Minh City $70000$, Vietnam}
\affiliation{Faculty of Natural Sciences, Duy Tan University, Da Nang City $50000$, Vietnam}
\author{T.T. Hong} \email{tthong@agu.edu.vn}
\affiliation{An Giang University, Long Xuyen City 880000, Vietnam} 
\affiliation{Vietnam National University, Ho Chi Minh City 700000, Vietnam} 
\author{T. Phong Nguyen}\email{thanhphong@ctu.edu.vn}
\affiliation{Department of Physics, Can Tho University,
	3/2 Street, Can Tho, Vietnam}
\author{N. H. T. Nha} \email{nguyenhuathanhnha@vlu.edu.vn}
\affiliation{Subatomic Physics Research Group, Science and Technology Advanced Institute, Van Lang University, Ho Chi Minh City, Vietnam}
\affiliation{Faculty of Applied Technology, School of Technology,  Van Lang University, Ho Chi Minh City, Vietnam}

\begin{abstract}
 General expressions for one-loop contributions associated with  lepton-flavor violating decays of the standard model-like Higgs boson $h\to  e_b^\pm e_a^\mp$ and gauge boson $Z\to e^\pm_b e_a^\mp$ are introduced in the unitary gauge.  The results are used to discuss these decays as new physics signals in a minimal left-right symmetric model containing only one bidoublet Higgs and a $SU(2)_R$ Higgs doublet accommodating data of neutrino oscillations and $(g-2)_{\mu}$. The numerical investigation indicates that some of these decay rates can reach near future experimental sensitivities.
%
\end{abstract}

\maketitle
\allowdisplaybreaks

\section{Introduction}
Lepton flavor violating (LFV) decays, like those of charged leptons (cLFV) $e_b\to e_a \gamma$,  the standard model-like Higgs boson  (LFV$h$) $h\to e_be_a$,  and the neutral gauge boson (LFV$Z$) $Z\to e_be_a$, are hot objects of experimental searches \cite{CMS:2021rsq, ATLAS:2019xlq,CMS:2023pte,ATLAS:2023mvd, ATLAS:2021bdj,ATLAS:2022uhq}. Although these decays do not appear in the standard model (SM), their existence is predicted by the LFV sources appearing in many models beyond the SM (BSM),  such as minimal  SM  extensions guaranteeing neutrino oscillation data \cite{Pilaftsis:1992st, Korner:1992zk}. Along with updates of the experimental data, LFV decays of $h$ and $Z$, as promising signals of new physics, were indicated in various BSMs. On the other hand, the recent experimental results of  charged lepton anomalies $ a_{e_a}\equiv (g-2)_{e_a}/2$ show a large deviation from the SM, which often supports regions of parameter space predicting large LFV decay rates, especially for the BSM accommodating neutrino oscillation data.  Therefore, a combination of simultaneous studies of the above LFV decays and the $(g-2)_{e_a}$ data in BSMs accommodating the neutrino oscillation data is beneficial for searching for allowed regions of the parameter space of the models. To the best of our knowledge,  in addition to the  analytical formulas for one-loop contributions to cLFV decays and $(g-2)_{e_a}$ anomalies \cite{Lavoura:2003xp, Crivellin:2018qmi}, which can generally be applied  to a large class of BSMs, the one-loop contributions to LFV$h$ and LFV$Z$ decay amplitudes were introduced in specific SM extensions. Namely, various discussions on these LFV decays of $h$   \cite{Diaz-Cruz:1999sns, Arganda:2004bz, Arganda:2014dta,  Hue:2015fbb, Arganda:2015uca, Arganda:2015naa, Baek:2015mea, Chiang:2016vgf, Thuc:2016qva, Arganda:2016zvc, Cai:2017jrq, Nguyen:2018rlb, Nguyen:2020ehj, Zhang:2021nzv,  Hong:2022xjg, CarcamoHernandez:2020pnh, Marcano:2019rmk, Chen:2023eof}  
 and $Z$
 \cite{Korner:1992an,Ilakovac:2012sh, DeRomeri:2016gum,  Crivellin:2018mqz, Herrero:2018luu, Jurciukonis:2021izn,  Hernandez-Tome:2019lkb,  Abada:2021zcm, Abada:2022asx, Crivellin:2022cve,  Hong:2023rhg, Calibbi:2021pyh, Hong:2024yhk, Hong:2024swk,  Kriewald:2022erk}  
 as signals of new physics originating from loop contributions.   On the other hand,  BSMs consisting of both heavy seesaw neutrinos and right-handed gauge bosons, such as the left-right (LR) symmetric models \cite{Pati:1974yy,Mohapatra:1974gc,Mohapatra:1974hk,Senjanovic:1975rk, Senjanovic:1978ev, Mohapatra:1979ia}, can predict complicated one-loop contributions to the LFV decay amplitudes.  Phenomenology of these LR models including the $(g-2)_{e,\mu}$ and cLFV decays  has been discussed recently \cite{ThomasArun:2021rwf, Khalil:2010iu, Ezzat:2021bzs, Ezzat:2022gpk, Ashry:2022maw}. 
  Therefore,  concrete studies of LFV$h$ and $Z$ decays will be more useful for determining the allowed regions of parameter space that are not excluded by experimental constraints. Our main aim is to introduce a general complete class of one-loop contributions for LFV decay amplitudes calculated in the unitary gauge. We will use these  results to discuss in detail the correlations among LFV decay rates and $(g-2)_{\mu}$ in  the LR model discussed in Ref. \cite{Ashry:2022maw}.

The paper is organized as follows. In section II, we review the one-loop contributions to decay amplitudes $e_b\to e_a \gamma$ and $(g-2)_{e,\mu}$  and provide general one-loop contributions to decay amplitudes $h\to e_a^\pm e_b^\mp$ and $Z\to e_a^\pm e_b^\mp$, which are derived using the unitary gauge. In section III, we determine analytic formulas for one-loop contributions to LFV decay rates, which are used to study the allowed regions of the parameter space that guarantee simultaneously the neutrino oscillation data,  LFV constraints, and $1\sigma$ deviation of $(g-2)_{\mu}$ data from the SM prediction.  Section IV summarizes important results. Finally,  three appendices are used to provide more detailed notations of Passarino-Veltman (PV) functions, precise expressions of relevant one-loop formulas, and the Higgs sector of the LRIS model.  

\section{General one-loop formulas}
\subsection{cLFV decays $e_b\to e_a \gamma$ and $(g-2)_{e_a}$}
\begin{figure}[ht] 
	\includegraphics[width=14cm]{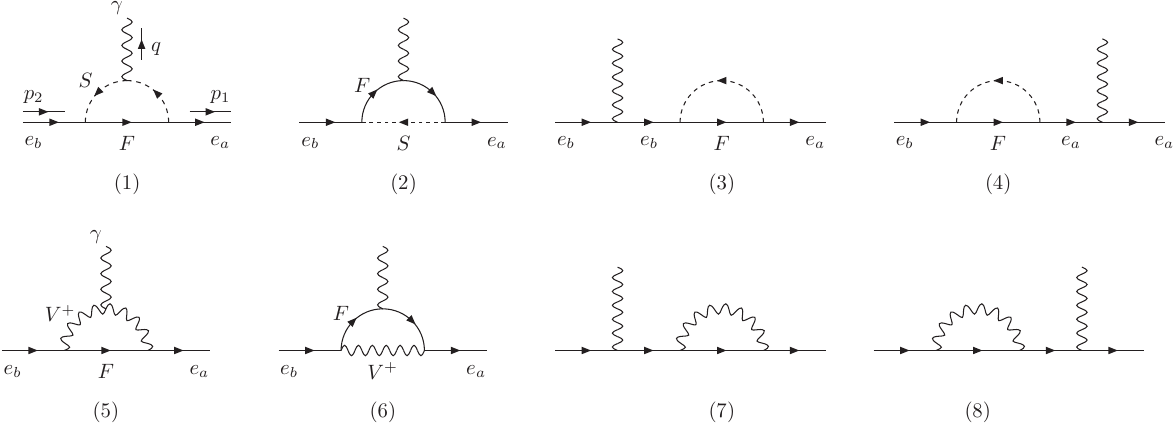}
	\caption{ One-loop contributions to the decay amplitudes $e_b\to e_a\gamma$ and $(g-2)_{e_a}$.}\label{fig_ebaU}
\end{figure}

The  one-loop contributions  to cLFV decay amplitudes and $a_{e_a}$ are  available \cite{Lavoura:2003xp, Crivellin:2018qmi, Hue:2017lak, Hue:2023rks}, where calculations were performed in both gauges  't Hooft-Feynman and unitary for diagrams consisting of gauge boson exchanges. We adopt the following Lagrangian parts playing the roles of LFV sources discussed in this work  \cite{Crivellin:2018qmi}
\begin{align}
\mathcal{L}_{FeS} &= \sum_{F,S}\sum_{a=1}^3 \overline{F}(g_{a FS}^{L}P_L +g_{aFS}^{R}P_R) e_a S +\mathrm{h.c.},	\label{eq_LFh}
\\
\mathcal{L}_{FeV}& =  \sum_{F,V}\sum_{a=1}^3  \overline{F}\gamma^{\mu}  (g_{aFV}^{L}P_L +g_{a FV}^{R}  P_R)e_aV_{\mu} +\mathrm{h.c.} \label{eq_LFV},
\end{align}
where the fermion $F$ and the  boson $B=V_{\mu},S$ have  electric charges $Q_{F}$ and $Q_{B}$,    and  masses $m_{F}$ and  $m_{B}$, respectively. This means that $\hat{Q}B=Q_B$ and $\hat{Q}B^*=-Q_B$, therefore the conventions $B\equiv B^{+Q_B}$ and $B^*\equiv B^{-Q_B}$ are used hereafter.  Two Eqs.  \eqref{eq_LFh} and \eqref{eq_LFV} are consistent with those introduced in Ref. \cite{Lavoura:2003xp}. Moreover,  we adopt the Feynman rule that the  photon always couples with two identical physical particles  \cite{Lavoura:2003xp}, as  shown in Table \ref{t_AXX},
\begin{table}[h]
	\begin{tabular}{|c|c|c|c|c|c|}
		\hline
		Vertex & Coupling & 	Vertex  & Couplings &Vertex  & Couplings\\
		\hline
		$A^{\mu}(p_0)V^{\nu}(p_+)V^{*\lambda}(p_-)$&$-ieQ_{V}\Gamma_{\mu \nu \lambda}(p_0,p_+,p_-) $&$A^{\mu}S(p_+)S^*(p_-)$&$ ieQ_{S}(p_+-p_-)_{\mu}$ & $A^{\mu}\overline{F}F$& $ieQ_{F}\gamma_{\mu}$\\
		\hline
	\end{tabular}
	\caption{Feynman rules for cubic couplings of photon $A^{\mu}$. 
		\label{t_AXX}}
\end{table} 
where $\Gamma_{\mu \nu \lambda}(p_0,p_+,p_-) = g_{\mu\nu} (p_0 -p_+)_{\lambda} +g_{\nu \lambda} (p_+ -p_-)_{\mu} +g_{ \lambda \mu} (p_- -p_0)_{\nu}$ is the standard form with $\partial_{\mu}=-ip_{\mu}$.  Here,  $p_{\mu}= p_{0,\pm}$ are incoming momenta into the relevant vertex consisting of a neutral,  and two charged conjugated gauge bosons $V^0\;(=A_{\mu},Z_{\mu})$, $B$, and $B^*$,  respectively. The Ward Identity forbids the tree-level couplings of a photon with two different physical states \cite{Hue:2023rks}.  In the unitary gauge, the one-loop contributions to the decay amplitudes $e_b\to e_a \gamma$ and $a_{e_a}$ are shown in Fig. \ref{fig_ebaU}. 

Previous calculations in the unitary  gauge were performed, including approximate formulas with heavy gauge boson exchanges \cite{Crivellin:2018qmi}, or exact  ones \cite{Hue:2023rks} expressed in terms of the PV functions \cite{tHooft:1978jhc}, which are convenient for computation using numerical packages such as  LoopTools  \cite{Hahn:1998yk}. They are also consistent with the calculation in the 't Hootf-Feynman (HF) gauge based on the particular assumption of the Goldstone boson couplings \cite{Lavoura:2003xp},  where the form factors are 
given in appendix \ref{app_LFVU}. Useful transformations between different notations used in previous works are given  in Ref. \cite{Hue:2023rks}. 

We note here that two Eqs. \eqref{eq_LFh} and \eqref{eq_LFV} contain  general LFV sources generating 1-loop contributions to  other LFV processes, including the LFV$h$ and LFV$Z$ we focus on in this work. The most attractive LFV source comes from the active neutrino oscillation data, the evidence of the LFV source confirmed by experiments \cite{Super-Kamiokande:1998kpq, Super-Kamiokande:2000ywb, SNO:2002tuh}. Especially the heavy neutrinos generating active neutrino masses and mixing through the seesaw mechanism have couplings  with forms given in two Eqs. \eqref{eq_LFh} and \eqref{eq_LFV}.  Usually, only the  cLFV experimental data of $\mu \to e \gamma$ are considered as the strictest constraints on the allowed regions of the parameter space, successfully explaining the $(g-2)_{e,\mu}$ data \cite{Li:2022zap, Hong:2023rhg, CarcamoHernandez:2024edi,  Escribano:2021css}.  Recent discussions on ISS extensions  of the 3-3-1 models suggest that the decay $ \tau \to \mu \gamma$ may  result in stricter constraints on $(g-2)_{\mu}$ data than that of the decay $\mu \to e\gamma$ \cite{Hue:2021xap, Hong:2022xjg, Hong:2023rhg}. This suggests that the regions of parameter space giving large one-loop contributions to $(g-2)_{\mu}$ may also be affected by the experimental data of LFV$h$ and LFV$Z$ decays. We will discuss this in detail in the LRIS model \cite{Ashry:2022maw}.

\subsection{The LFV decays  $Z\to e_b^\pm  e_a^\mp$ and $h\to e_b^\pm  e_a^\mp$}

Unlike photon couplings, the gauge boson $Z$ and neutral SM-like Higgs boson $h$ can couple two different physical particles. In particular, the triple couplings of $Z$ relating to one-loop contributions to the decay amplitude $Z\to e^\pm_be^\mp_a$ are generally given  in Table \ref{t_ZXY}.  
\begin{table}[h]
	\begin{tabular}{|c|c|c|c|}
	\hline
	Vertex & Coupling & 	Vertex  & Couplings\\
	\hline
	$Z^{\mu}(p_0)V^{\nu}(p_+)V'^{*\lambda}(p_-)$&$-ieg_{ZVV'}\Gamma_{\mu \nu \lambda}(p_0,p_+,p_-) $&$Z^{\mu} S'^*(p_-)S(p_+)$&$ ieg_{ZS'^* S}(p_+-p_-)_{\mu}$ \\
	\hline
	$SV^{\mu*}Z^{\nu}$&$ ieg_{SZV}g_{\mu\nu}$ &$S^{*}V^{\mu}Z^{\nu}$&$ ieg_{SZV}^* g_{\mu\nu}$ \\
	\hline
	$Z^{\mu}\overline{F}F'$&$ie\gamma_{\mu}\left(  g_{ZFF'}^LP_L +g_{ZFF'}^RP_R \right)$ & $Z^{\mu}\overline{F'}F$&$ie\gamma_{\mu}\left( g_{ZFF'}^{L*}P_L +g_{ZFF'}^{R*}P_R \right)$ \\
	\hline
\end{tabular}
	\caption{Feynman rules for cubic couplings of  $Z^{\mu}$ with conventions defined in Table \ref{t_AXX}.   
\label{t_ZXY}}
\end{table}
 The triple self-couplings of $Z$  are included in the kinetic parts of gauge bosons. The couplings of $Z$ with fermions are included in the fermion kinetic Lagrangian. The triple couplings of $Z$ with two bosons are included in the covariant kinetic Lagrangian of scalar multiplets $R_S$,  $\mathcal{L}^S_{\mathrm{kin}}  =  \left(D_{\mu} R_{S}\right)^{\dagger} \left(D^{\mu} R_{S}\right)$, where  $D_{\mu} =\partial_{\mu}- iP_{\mu}$.  If we denote  that $(\partial_{\mu}R_{S}^{\dagger}) P^{\mu}R_S\equiv \sum_{V, S',S} g_{VS^*S'} (\partial_{\mu}S^{*})  V^{\mu}S'$, then it results in the following part for  $V-S-S'$ couplings
 \begin{align}
 	\label{eq:LVSS'}
 	\mathcal{L}_{VSS'}  =& \sum_{V,S,S'} ig_{VS^*S'}\left[-(\partial_{\mu}S^*)S' +S^*(\partial_{\mu}S')  \right]V^{\mu} +\mathrm{H.c.}
 	\crn =& \sum_{V,S,S'} \left[ g_{VS^*S'} (p_{S' \mu} -p_{S^*\mu}) S^*S'V^{\mu} + g^*_{VS^*S'} (p_{S\mu} -p_{S'^* \mu}) SS'^*V^{*\mu}\right],
 \end{align}
 where the rules  used to transform into the momentum space are $ (\partial_{\mu}B)=-ip_{B\mu}B$ with  $B=S,S',S^*,S'^*$, and all momenta are incoming vertex conventionally.  The LFV$Z$ and LFV$h$ decays have couplings $ZS_kS_l$ and $VSh$, respectively. The notations show  that $g_{VS^*S'}^*=g_{VS'^*S}$. The couplings in the second row of Table \ref{t_ZXY} are derived from the covariant part containing at least one neutral Higgs component with non-zero vacuum expectation values (vev), namely $P_{\mu} R_S\to V_{\mu } \langle S^0\rangle + V'_{\mu} S$. Such couplings do not contain momentum, hence it is easy to determine that 
 $$	\mathcal{L}_{SVV'}=\sum_{S,V,V'} g_{SVV'} g_{\mu\nu} SV^{\mu}V'^{*\nu} + \mathrm{H.c.},$$
 which corresponds  to the Feynman rules in the second row of table \ref{t_ZXY} for $Z=V'\neq V$. We will identify $V'^{\mu}=W^{+\mu}$ for consistency with the SM notation when studying certain BSMs. 

\begin{figure}[ht] 
	\includegraphics[width=14cm]{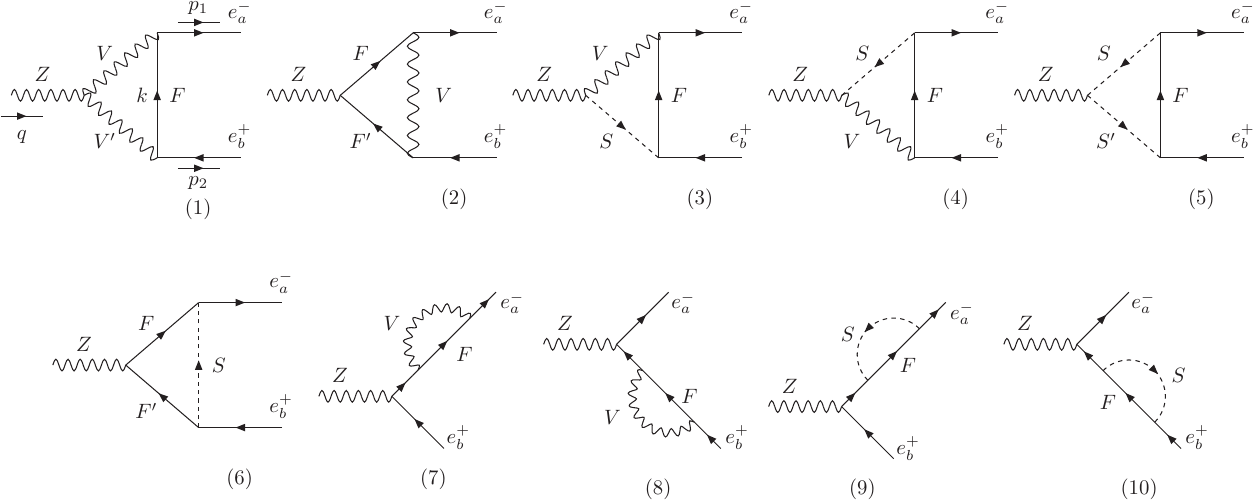}
	\caption{ One-loop contributions to the decay amplitudes $Z\to e_b^+ e_a^-$ in the unitary gauge.}\label{fig_ZebaG}
\end{figure}

The one-loop Feynman diagrams for LFV$Z$ decays are shown in Fig. \ref{fig_ZebaG}. 
We focus only on the unitary gauge so that  particles in all diagrams are physical.  The effective amplitude for the LFV$Z$ decays  $Z\to e^+_b (p_2) e^{-}_a (p_1)$ is written following the notations \cite{Jurciukonis:2021izn, DeRomeri:2016gum,Hong:2023rhg}:
\begin{align} \label{eq_Mzeab}
	i\mathcal{M}(Z\to e^+_be^-_a)
	%
	= & \frac{ie}{16\pi^2} \overline{u}_{a}\left[ \slashed{\varepsilon} \left( \bar{a}_l P_L + \bar{a}_r P_R\right) +  (p_1.\varepsilon) \left( \bar{b}_l P_L + \bar{b}_r P_R\right)  \right]v_{b},
\end{align}
where $\varepsilon_{\alpha}(q)$ is  the polarization of $Z$ and $u_a(p_1)$, and $v_b(p_2)$ are  Dirac spinors of $e_a^-$ and $e^+_b$. All form factors $\bar{a}_{l,r}$ and $\bar{b}_{l,r}$  receive contributions from one-loop corrections. The external on-shell gauge boson $Z$ gives  $q.\varepsilon=0$, leading to $p_2.\varepsilon=-p_1.\varepsilon$. The on-shell conditions of the final leptons and $Z$ boson are $p_1^2= m_1^2 =m_a^2$, $p_2^2=m_2^2 =m_b^2$, and $q^2=m_Z^2$. The respective partial decay width is 
\begin{align} 
	\label{eq_GAZeba}
	\Gamma (Z\to e^+_b e^-_a)= 	\frac{\sqrt{\lambda}}{16\pi m_Z^3}\times \left(\frac{e}{16\pi^2}\right)^2 \left( \frac{\lambda M_0}{12 m^2_Z} +M_1 +\frac{ M_2}{3 m^2_Z}\right),
\end{align}
where $\lambda= m^4_Z +m^4_{b} +m^4_{a} -2(m^2_Zm^2_{a} +m^2_Zm^2_{b} +m^2_{a}m^2_{b})$, 
and 
\begin{align}
	\label{eq_Mi}
	M_0= & (m^2_Z -m_{a}^2 -m_{b}^2)\left(|\bar{b}_l|^2 +|\bar{b}_r|^2\right)  -4 m_{a} m_{b} \mathrm{Re}\left[ \bar{b}_l  \bar{b}^*_r\right]
	\crn&
	- 4m_{b} \mathrm{Re}\left[ \bar{a}^*_r \bar{b}_l   + \bar{a}^*_l \bar{b}_r  \right] -  4m_{a}\mathrm{Re}\left[ \bar{a}^*_l \bar{b}_l   + \bar{a}^*_r  \bar{b}_r  \right] , 
	\crn M_1 = & 4 m_{a}m_{b} \mathrm{Re}\left[\bar{a}_l\bar{a}_r^* \right],
	\crn  M_2 = &  \left[ 2 m^4_Z - m_Z^2\left( m_{a}^2 + m_{b}^2\right) - \left( m_{a}^2 - m_{b}^2\right)^2  \right] \left( |\bar{a}_l|^2 +|\bar{a}_r|^2\right).
\end{align}
The total form factors consist of all the particular one-loop contributions originating from the diagrams given in Fig. \ref{fig_ZebaG}.  They are divided into three parts with different virtual particle exchanges in the loops: pure gauge bosons, pure scalars, and the appearance of both gauge bosons and scalars. The respective contributions are listed  as follows
\begin{align}
	\label{eq_abZeba}
	\bar{a}_{L,R}&= \bar{a}^{(1+2+7+8)}_{L,R} +\bar{a}^{(5+6+9+10)}_{L,R} +\bar{a}^{(3+4)}_{L,R},  
	\crn  \bar{b}_{L,R} &= \bar{b}^{(1+2+7+8)}_{L,R} +\bar{b}^{(5+6+9+10)}_{L,R} +\bar{b}^{(3+4)}_{L,R}, 
	\crn 	\bar{a}^{(1+2+7+8)}_{L,R} & = \sum_{V,V',F}\bar{a}^{FVV'}_{L,R}+ \sum_{V,F,F'}\bar{a}^{VFF'}_{L,R} +  \sum_{V,F}\bar{a}^{FV}_{L,R},
	\crn 	\bar{b}^{(1+2+7+8)}_{L,R} & = \sum_{V,V',F}\bar{b}^{FVV'}_{L,R}+ \sum_{V,F,F'}  \bar{b}^{VFF'}_{L,R} +\sum_{V,F} \bar{b}^{FV}_{L,R},
	\crn 	\bar{a}^{(5+6+9+10)}_{L,R} & =  \sum_{S,S',F}\bar{a}^{FSS'}_{L,R}+ \sum_{S,F,F'}\bar{a}^{SFF'}_{L,R}  +\sum_{S,F} \bar{a}^{FS}_{L,R},
	\crn 	\bar{b}^{(5+6+9+10)}_{L,R} & =  \sum_{S,S',F}\bar{b}^{FSS'}_{L,R}+ \sum_{S,F,F'}\bar{b}^{SFF'}_{L,R}  +\sum_{S,F} \bar{b}^{FS}_{L,R},
	\crn \bar{a}^{(3+4)}_{L,R} &= \sum_{S,V,F} \left( \bar{a}^{FSV}_{L,R} +\bar{a}^{FVS}_{L,R}\right),
	\crn \bar{b}^{ (3+4)}_{L,R} &= \sum_{S,V,F} \left( \bar{b}^{FSV}_{L,R} +\bar{b}^{FVS}_{L,R}\right),
\end{align}
where we omit the index $(ab)$ for simplicity, namely $\Delta^{ X}_{L,R}\equiv \Delta^{(ab) X}_{L,R}$ for all $X= FVV',\dots$. The analytical formulae of the form factors given in Eq. \eqref{eq_abZeba} were calculated by hand in the unitary gauge and cross-checked via the form package \cite{Vermaseren:2000nd}. The results are collected in appendix \ref{app_LFVU}, where all one-loop contributions are introduced in terms of the PV-functions consistent with those defined in LoopTools \cite{Hahn:1998yk}.  They will be used for the specific discussions in the LRIS model framework presented in section \ref{sec_LRIScontent}.  Apart from that, the analytic forms of the FV-functions in the limit $m_a=m_b=0$ were introduced in Ref. \cite{Hue:2015fbb},  where it was shown that they could be used approximately without using LoopTools \cite{Phan:2016ouz}.  

\begin{figure}[ht] 
	\includegraphics[width=14cm]{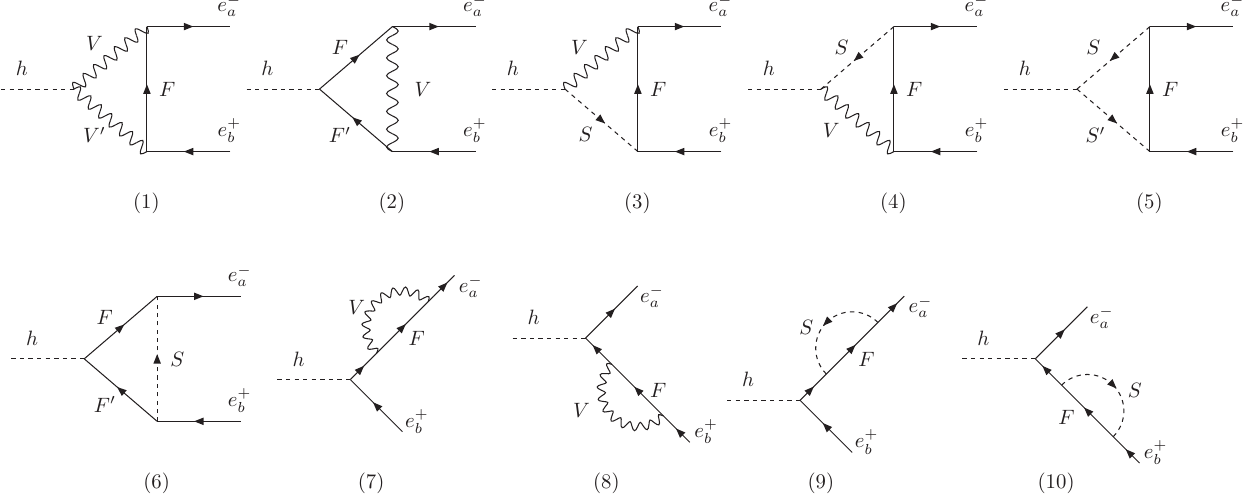}
	\caption{ One-loop contributions to the decay amplitudes $h\to e_b^+ e_a^-$ in the unitary gauge.}\label{fig_hebaG}
\end{figure}

The couplings of the SM-like Higgs boson $h$ related to one-loop contributions to the LFV$h$ decays are generally given  in Table \ref{t_hXY}.  
\begin{table}[h]
	\begin{tabular}{|c|c|c|c|}
		\hline
		Vertex & Coupling & 	Vertex  & Couplings\\
		\hline
		$hV^{\mu}V'^{*\nu}$&$ig_{hVV'} g_{\mu\nu}$ &$hSS'^*$&  $-i\lambda_{hSS'}$\\
		\hline
$V^{\mu }S^*(p_{-})h(p_0)$&$ ig_{VSh}(p_0 -p_{-})_{\mu}$ & $V^{*\mu }S(p_{+})h(p_0) $&$  ig^*_{VSh}(p_{+} -p_0 )_{\mu}$ \\
\hline
$h \overline{F}F'$&$- i \left( g_{hFF'}^LP_L +g_{hFF'}^RP_R \right)$ & 	$h \overline{F'}F$&$-i \left( g_{hFF'}^{L*}P_R +g_{hFF'}^{R*}P_L \right)$  \\
\hline
	\end{tabular}
	\caption{Feynman rules for cubic couplings of the SM-like Higgs boson $h$.   
		\label{t_hXY}}
\end{table}
Although the general one-loop formulas related to gauge boson exchanges have been determined in many works,  the diagrams arising from the $hSV$ couplings were first discussed in Ref. \cite{Hue:2015fbb}, for a 3-3-1 model including simple analytical forms used for numerical evaluations without the need for numerical packages such as LoopTools.  These diagrams were also mentioned in the 2HDM, along with studying the cLFV and LFV$Z$ decays, with or without $(g-2)_{e_a}$ anomalies \cite{Nguyen:2020ehj, Hong:2023rhg}. A class of general one-loop analytic formulas for LFV$h$ decay amplitudes was given in ref.  \cite{Zeleny-Mora:2021tym}, which are applicable only to  the 't Hooft-Veltman gauge. On the other hand, the general LFV gauge couplings with non-zero $g^R_{aFV} \neq 0$ given in Eq. \eqref{eq_LFV} have not been mentioned before.   The one-loop Feynman diagrams for LFV$h$ decays are shown in Fig. \ref{fig_hebaG}. 
The effective Lagrangian of the LFV$h$ decay  $h \rightarrow e_a^{\pm}e_b^{\mp}$ is
$$ \mathcal{L}^{\mathrm{LFV}h}= h \left(\Delta^{(ab)}_{L} \overline{e_a}P_L e_b +\Delta^{(ab)}_{R} \overline{e_a}P_R e_b\right) + \mathrm{H.c.},$$
where $\Delta^{(ab)}_{L,R}$  arise from loop contributions.   
The respective partial decay width in the limit  $m_{h}\gg m_{a,b}$  is   \cite{Arganda:2004bz}
\begin{equation}
	\Gamma (h \rightarrow e_ae_b)\equiv\Gamma (h\rightarrow e_a^{-} e_b^{+})+\Gamma (h \rightarrow e_a^{+} e_b^{-})
	\simeq   \fr{ m_{h}}{8\pi}\left(\vert \Delta^{(ab)}_L\vert^2+\vert \Delta^{(ab)}_R\vert^2\right). \label{eq_LFVwidth}
\end{equation}
 The corresponding branching ratio is  Br$(h\rightarrow e_ae_b)= \Gamma (h\rightarrow e_ae_b)/\Gamma^{\mathrm{total}}_{h}$ where $\Gamma^{\mathrm{total}}_{h}\simeq 4.1\times 10^{-3}$ GeV \cite{LHCHiggsCrossSectionWorkingGroup:2016ypw} with  $ q^2 \equiv( p_1+p_2)^2=m^2_{h}$.  Similar to the case of discussion on the LFV$Z$ decays,   the  particular formulas  one-loop contributions to   $\Delta^{(ab)}_{L,R}$ are given in appendix \ref{app_LFVU}, namely
\begin{align}
	\label{eq_DeLRh}
	\Delta^{(ab)}_{L,R} &=\Delta^{(1+2+7+8)}_{L,R} +\Delta^{(5+6+9+10)}_{L,R} +\Delta^{(3+4)}_{L,R}, 
	\crn \Delta^{(1+2+7+8)}_{L,R} &= \sum_{F,V,V'}\Delta^{ FVV'}_{L,R} + \sum_{F,F',V} \Delta^{VFF'}_{L,R}   + \sum_{F,V}\Delta^{FV}_{L,R},
	\crn \Delta^{(5+6+9+10)}_{L,R} &= \sum_{F,S,S'}\Delta^{FSS'}_{L,R} +\sum_{F,F',S} \Delta^{ SFF'}_{L,R} + \sum_{F,S} \Delta^{ FS}_{L,R},
	\crn \Delta^{(3+4)}_{L,R} &= \sum_{F,S,V} \left( \Delta^{FSV}_{L,R} +\Delta^{ FVS}_{L,R}\right),
\end{align} 
where we omit the index $(ab)$ for simplicity, namely $\Delta^{ X}_{L,R}\equiv \Delta^{(ab) X}_{L,R}$ for all $X= FVV',\dots$.
We note here important properties of the neutral boson couplings to fermions, namely, the two vertices $h\bar{F}F'$ and $Z\bar{F}F'$ may have different vertex factors from the respective Lagrangian parts. On the basis of the general Feynman rules for four-component spinors of   fermions   discussed in Ref. \cite{Dreiner:2008tw}, two classes of the Lagrangian parts corresponding to the appearance of the Dirac or Majorana spinors are considered here. In particular, the Lagrangian for at least one Dirac spinor is as follows:
\begin{align}
	\label{eq_ZhfD}
\mathcal{L}^D_{Zff} =&e \sum_{F,F'} \left[  \overline{F} \gamma_{\mu} \left(  g_{ZFF'}^LP_L +g_{ZFF'}^{R}P_R \right) F' Z^\mu +\mathrm{h.c.}\right] ,
\crn \mathcal{L}^D_{hff} =  &  -\sum_{F,F'} \left[ \overline{F}\left(  g_{hFF'}^LP_L +g_{hFF'}^{R}P_R \right) F'h +\mathrm{h.c.} \right].  
\end{align}
 Various available  BSMs consist of Dirac fermions  and scalars to accommodate the $(g-2)_{\mu}$ anomaly  data, such as vector-like leptons \cite{Crivellin:2018qmi, Ghorbani:2022muk} and SM quarks in leptoquark models \cite{Bhaskar:2022vgk}.   Further studies to explain successfully simultaneous $(g-2)_{e}$ data with minimal Dirac fermion couplings with different charged leptons may result in promising  LFV$h$ and LFV$Z$ decay signals. 

In contrast, the Lagrangian for  vertices consisting of two Majorana spinors $F=\left(F\right)^c$ and $F' =\left(F'\right)^c$  is
\begin{align}
	\label{eq_ZhfM}
	\mathcal{L}^M_{Zff} =&  \frac{e}{2}  \sum_{F,F'}\overline{F} \gamma_{\mu} \left(  g_{ZFF'}^LP_L -g_{ZFF'}^{L*}P_R \right) F' Z^\mu,
	\crn \mathcal{L}^M_{hff} =  &  - \frac{1}{2} \sum_{F,F'} \overline{F}\left(  g_{hFF'}^LP_L +g_{hFF'}^{L*}P_R \right) F'h,
\end{align}
where the Majorana conditions give $g_{hFF'}^{R}=g_{hFF'}^{L*}$ and $g_{ZFF'}^{R}=-g_{ZFF'}^{L*} =-g_{ZF'F}^{L}$. The usual conventions  were used previously  $g_{ZF_{i}F_{j}}^{L} \varpropto  q_{ij}, C_{ij}$ in Refs. \cite{Jurciukonis:2021izn, Arganda:2014dta} and $g_{hF_{i}F_{j}}^{L} \varpropto \lambda^h_{ij},\lambda_{ij}$ \cite{Arganda:2014dta}, for example. These works considered $F,F'\equiv n_i,n_j$ as Majorana neutrinos explaining the experimental neutrino oscillation data through the standard seesaw (SS) or inverse seesaw (ISS) mechanism.  After  the Feynman rules are constructed,  the same calculations are applicable  to both Dirac and Majorana spinors using the conventions for four-component spinors \cite{Dreiner:2008tw}.   

At tree level, we adopt a Lagrangian  for the  SM-like couplings of $Z$ and $h$ to SM leptons  $Z\overline{e_b}e_a$ and $h\overline{e_b} e_a$ in this model is:
\begin{align}
	\label{eq_Lnn}
	\mathcal{L}_{nc}&= e Z_{\mu} \sum_{a=1}^3  \bar{e}_a \gamma^{\mu}\left[t_L P_L +t_R  P_R\right] e_a -\frac{gm_{a}}{2m_W}\delta_{hee} h\overline{e_a} e_a, 
\end{align}
in which  the SM limit  results in the following relations  
\begin{equation}\label{eq_qịj}
	t_R=t^{\mathrm{SM}}_R=\frac{s_W}{c_W},\; t_L=t^{\mathrm{SM}}_L=\frac{s^2_W-c^2_W}{2s_Wc_W}, \; \delta_{hee}=1. 
\end{equation}

\section{ \label{sec_LRIScontent} The LRIS model and Feynman rules for LFV decays}
\subsection{Brief review of the LRIS model } 
The LR symmetry models constructed based on the gauge group $SU(3)_c \times SU(2)_L\times SU(2)_R\times U(1)_{B-L}$ have been widely studied, such as the minimal model \cite{Dev:2016dja} with a minimal seesaw (MSS) or linear seesaw (LSS) mechanism to generate neutrino mass and mixing. Another extension of the LR  version (LRIS)  was introduced \cite{Khalil:2010iu, Ezzat:2021bzs, Ezzat:2022gpk} with inverse seesaw (ISS) neutrinos  to successfully explain the $R_D$ and $R_{D^*}$ anomalies with a rather light scalar spectrum. The $(g-2)_{e,\mu}$ anomalies were also discussed in LRIS models \cite{Ashry:2022maw} and the correlations with the cLFV decay Br$(\mu \to e\gamma)$.  It is therefore interesting to study other LFV$Z$ and LFV$h$ decays  in this model. We summarize here the particle content and Lagrangian discussed in the above works.  
The charged operator is defined as 
\begin{equation}
Q=T^L_3 + T^R_3+\dfrac{B-L}{2}=T^{L}_3+\dfrac{Y}{2},
\label{eq:charge_Q}
\end{equation}
where  $T^{L,R}_3$ are  the generators of $SU(2)_{L,R}$, and $B$ and $L$ are   the baryon and lepton numbers of the $U(1)_{B-L}$ group.  The matching condition with SM gives  $\dfrac{Y}{2}=T^R_3+\dfrac{B-L}{2}$.

The particle content bases corresponding to the  gauge group $SU(3)_c\times  SU(2)_L\times SU(2)_R\times U(1)_{B-L}\times Z_2$ is:
 \begin{align}
 \label{eq_LRparticle}
 	Q_{L}= & \begin{pmatrix}
 	u	\\
 	d	
 	\end{pmatrix}_{L} \sim \left(3,2,1,\frac{1}{3},+\right),   \;  L_{L}=  \begin{pmatrix}
 	\nu	\\
 	e	
 	\end{pmatrix}_{L} \sim \left(1,2,1,-1,+\right), 
\crn 	Q_{R}= & \begin{pmatrix}
	u	\\
	d	
\end{pmatrix}_{R} \sim \left(3,1,2,\frac{1}{3},+\right),   \;  L_{R}=  \begin{pmatrix}
	\nu	\\
	e	
\end{pmatrix}_{R} \sim \left(1,1,2,-1,+\right), 
\crn  	S_1 &\sim \left(1,1,1,0,-\right),  \; 	S_2 \sim \left(1,1,1,0,+\right),
\crn \phi &= \begin{pmatrix}
\phi^0_1	&  \phi^+_1\\
\phi^-_2	&  \phi^0_2
\end{pmatrix} \sim \left( 1,2,2,0,+\right), \; \chi_R =   \begin{pmatrix}
 \chi_R^+ \\
 \chi_R^0
\end{pmatrix} \sim \left( 1,1,2,1,+\right).
 \end{align}
 The $U(1)_{B-L}$ charges of $S_{1,2}$ are consistent with ref. \cite{ThomasArun:2021rwf},  guaranteeing the  zero electric charges resulting from Eq. \eqref{eq:charge_Q}. The neutral Higgs components  get the following non-zero vev:
 \begin{equation}
 	\label{eq_vev}
 	\langle \phi\rangle =\mathrm{diag} \left( \frac{v_1}{\sqrt{2}},\; \frac{v_2}{\sqrt{2}}\right), \; \langle \chi^0_R\rangle =\frac{v_R}{\sqrt{2}}. 
 \end{equation}
The parameter $t_{\beta}\equiv v_1/v_2$ and $v^2\equiv v_1^2+v_2^2$ will be used when matching with the SM.  We focus on the Yukawa part of leptons as follows:
 \begin{align}
 	\label{eq_YL}
 	- \mathcal{L}_Y= & \sum_{i,j=1}^3 \left[ y^L_{ij}\overline{L_{R_i}}\phi^{\dagger} L_{Lj} + \tilde{y}^L_{ij}\overline{L_{R_i}}\tilde{\phi}^{\dagger} L_{Lj} + y^s_{ij}\overline{L_{R_i}} \tilde{\chi}_R(S_{2j})^c+\mathrm{H.c.}\right],
 \end{align}
 where $\tilde{\phi}=  \sigma_2\phi^*\sigma_2$, and $\tilde{\chi}_R=i\sigma_2 \chi^*$.  
This generates the charged lepton mass matrix $\mathcal{M}^\ell =(y^Lc_{\beta} +\tilde{y}^L s_{\beta}) v/\sqrt{2}$, and the following neutrino mass matrix after symmetry breaking:
\begin{align}
	-\mathcal{L}^{\nu}_{mass} =&  \overline{\nu_R} m_D\nu_L + \overline{\nu_R}M_R^T (S_2)^c + \frac{\mu_s}{2} \overline{ S_2}(S_2)^c +\mathrm{h.c.}
\crn= & \frac{1}{2} \left( \overline{(\nu_L)^c},\; \overline{\nu_R}, \overline{S_2}\right) \mathcal{M}^{\nu}\left( \nu_L,\; (\nu_R)^c, (S_2)^c \right)^T+\mathrm{h.c.},
\end{align}
where $\mathcal{M}^{\nu}$ is a symmetric $9\times 9$ matrix having the following ISS form 
\begin{align}
\label{eq:Mnu}
\mathcal{M}^{\nu}  &= \begin{pmatrix}
		\mathcal{O}_{3\times 3}& m^T_D & \mathcal{O}_{3\times 3}\\
		m_D&  \mathcal{O}_{3\times3}& M_R^T\\
		\mathcal{O}_{3 \times3}&M_R& \mu_s
	\end{pmatrix}, \; m_D= \frac{v}{\sqrt{2}} \left( y^Ls_{\beta} +\tilde{y}^L c_{\beta}\right), \; M_R = \frac{y^{sT}}{\sqrt{2}}v_R, 
\end{align}
where 
$\nu_L=(\nu_1, \nu_2, \nu_3)_L^T, \; \nu_R=(\nu_1,\nu_2, \nu_3)_R^T, \; S_2=(S_{21},S_{22},S_{23})^T.$  The total mixing matrix is defined as a $9\times 9$ unitary  matrix $U^{\nu}$ satisfying
\begin{align}
	\label{eq_Unu}
	U^{\nu T} 	\mathcal{M}^{\nu} U^{\nu } &= \hat{\mathcal{M}}^{\nu}=\mathrm{diag}(m_{n_1},\;m_{n_2},\;...,m_{n_{9}})= \mathrm{diag}(\hat{m}_{\nu},\hat{M}_N), 
	\crn n'_L &= U^{\nu} n_{L},\; 
n'_R= U^{\nu*} n_{R}= U^{\nu*} (n_{L})^c,\; 
\end{align} 
where the two left- and right-handed flavor base are  $n'_L=(	\nu_L, 		(\nu_R)^c, (S_2)^c )^T$ , and $(n'_L)^c=n'_R=(	\left(\nu_L\right)^c, \nu_R, S_2 )^T$,   $ n_{L,R}=(n_{1},n_{2},..., n_{9})_{L,R}$ are Majorana neutrino mass eigenstates  $n_{iL,R}=(n_{iR,L})^c$.  We will use the  approximate form  of $U^{\nu}$  as follows   \cite{Hong:2022xjg, Hong:2024yhk}
\begin{align}
	\label{eq:Usimeq}
	U^{\nu}&\simeq  \begin{pmatrix}
		\left(I_3 -\frac{R_0R_0^{\dagger}}{2}\right)U^{\nu}_3	& \frac{i R_0 }{\sqrt{2}}&  \frac{R_0}{\sqrt{2}}\\
		\mathcal{O}_{3\times3}	&  -\frac{iI_3}{\sqrt{2}}  &   \frac{I_3}{\sqrt{2}} \\
		-R_0^{\dagger} U^{\nu}_3& 	\frac{i}{\sqrt{2}}\left(I_3 -\frac{R_0^{\dagger}R_0}{2}\right)  &\frac{1}{\sqrt{2}}  \left(I_3 -\frac{R_0^{\dagger}R_0}{2}\right) 
	\end{pmatrix},
\end{align}
where 
\begin{align}
	m_D&= x_0^{\frac{1}{2}}  M_R \hat{\mu}_s^{-\frac{1}{2}}\xi\hat{x}^{\frac{1}{2}}_{\nu}U^{\nu \dagger}_3, \label{eq:mD22}
	\\ R_{0} &\equiv x_0^{\frac{1}{2}} U^{\nu }_3 \hat{x}^{\frac{1}{2}}_{\nu}\xi^{\dagger}\hat{\mu}_s^{-\frac{1}{2}} \label{eq:R0}, 
\\ M_R& =\hat{M}_R= \mathrm{diag}(M_1,M_2,M_3), \; \hat{M}_N=\mathrm{diag}(\hat{M}_R,\hat{M}_R),  \label{eq:hatMR}
\end{align}
and new conventions are 
\begin{align}
	\label{eq:x0}
	x_0 &\equiv \frac{\mathrm{max}[m_{n_1},m_{n_2},m_{n_3}]}{|(\mu_s)_{22}|}\ll 1, 
	\crn  \hat{\mu}_s & \equiv \frac{\mu_s}{|(\mu_s)_{22}|},\; \hat{x}_{\nu} \equiv \frac{ \hat{m}_{\nu}}{\mathrm{max}[m_{n_1},m_{n_2},m_{n_3}]} . 
\end{align}
The ISS condition $|\hat{m}_{\nu}|\ll |\mu_s|\ll |m_D|\ll M_{1,2,3}$  gives  $x_0\ll1$ but non-zero. Note that can be considered as the non-unitary scale of the active neutrino mixing matrix. 

Regarding the charged leptons, in general there are  two left and right rotations $U^{\ell}_{L,R}$ diogonalize the lepton mass matrix $\mathcal{M}^\ell$:
\begin{align}
\label{eq:VellLR}
U^{\ell \dagger}_R \mathcal{M}^\ell U^{\ell }_L = \hat{\mathcal{M}}^\ell=  \mathrm{diag}(m_e,m_{\mu},m_{\tau}),\; e_{L,R}\to U^\ell_{L,R} e_{L,R}. 
\end{align}
The  Pontecorvo-MakiNakagawa-Sakata (PMNS) matrix $U_{\mathrm{PMNS}}$   relating to the neutrino oscillation data is defined as $U_{\mathrm{PMNS}}=U^{\ell \dagger}_L U^{\nu}_3$ \cite{Maki:1962mu, Pontecorvo:1967fh, Workman:2022ynf}. It can be seen that \cite{Ashry:2022maw}:
\begin{align}
\label{eq:yL}
y^L=\frac{\sqrt{2}}{v c_{2\beta}} \left( \mathcal{M}^\ell c_{\beta} -m_D s_{\beta}\right),\; \tilde{y}^L= -\frac{\sqrt{2}}{v c_{2\beta}} \left( \mathcal{M}^\ell s_{\beta} -m_D c_{\beta}\right).  
\end{align}

The  covariant derivative  corresponding to bidoublet $\phi$ and doublets of the $SU(2)_{L,R}$ in the LRIS model are   \cite{Zhang:2007da}
%
\begin{itemize}
	\item For the  bidoublet such as $\phi$:
	\begin{align}
		\label{eq:Dmu22}
		D_\mu\phi&=\partial_\mu\phi  + \sum_{a=1}^3 \left[  -ig_L\dfrac{\sigma^a}{2}W^a_{L\nu}\phi+ig_R\phi\dfrac{\sigma^a}{2}W^a_R \right],
	\end{align}
	where $\sigma^a$ is the Pauli matrix.
	\item For  $SU(2)_{L,R}$ doublets such as $X_{L,R}=\chi_R,L_{iL,R},Q_{iL,R}$:
		\begin{align}
		\label{eq:Dmu2}
		D_\mu X_{A}&=\partial_\mu X_{A} - \sum_{a=1}^3 ig_A\dfrac{\sigma^a}{2}W^a_{A\mu} X_{A} - i g_{BL} B'_{\mu}\frac{B-L}{2} X_{A}, \; A=L,R. 
	\end{align}
\end{itemize}
Consequently, the kinetic Lagrangian of Higgs multiplets generating gauge boson masses are:
\begin{align}
\label{eq:LkH}
\mathcal{L}^H_k=\mathrm{Tr}\left[ (D_{\mu}\phi)^{\dagger} (D^{\mu}\phi)\right] +\left(D_{\mu}\chi\right)^{\dagger}\left(D^{\mu}\chi\right).
\end{align}
Defining $W^\pm_{ L,R \mu}\equiv (W^1_{ L,R \mu} \mp i W^2_{ L,R \mu})/\sqrt{2}$, the mixing angle of singly charged bosons $W-W'$ is $\theta$ determined as $t_{2\theta}=2s_{2\beta}v^2/v_R^2$, leading to the following relations of $W^\pm_{L,R \mu}$ and physical states $W_{\mu}$ and $W'_{\mu}$ \cite{Ezzat:2021bzs}
\begin{align}
\label{eq:thtaWLRpm}
\begin{pmatrix}
	W^\pm_{L\mu}	\\
	W^\pm_{R\mu}	
\end{pmatrix} =\left(
\begin{array}{cc}
	c_{\theta } & -s_{\theta } \\
	s_{\theta } & c_{\theta } \\
\end{array}
\right) \begin{pmatrix}
	W^\pm_{\mu}	\\
	W'^\pm_{\mu}	
\end{pmatrix}, 
\end{align}
 where we consider the simple case of $g_L=g_R=g_2=e/s_W$ is the $SU(2)_L$ gauge coupling of the SM.  In addition, $W$ is identified with the SM charged gauge boson with mass $m_W \simeq gv/2$,  where $v^2=v_1^2+v_2^2 =(246 \; \mathrm{GeV})^2$. The exact formulas of these two gauge boson masses are 
 \begin{align}
 \label{eq:mWx2}
 	m_W^2 &= \frac{g^2v^2}{4} \times \left( 1-t_{\theta}s_{2\beta}\right),\; m_{W'}^2= \frac{g^2v_R^2}{4} \times \frac{c_{\theta}^2 \left( s_{2\beta}+ t_{\theta}\right)}{c_{2\theta}s_{2\beta}},
 \end{align}
%
which result in the specific form of $\delta_{hee}^2= 1-t_{\theta}s_{2\beta}$, where $\delta_{hee}$ was  defined in Eq. \eqref{eq_Lnn}. 
 
 The mixing parameters and masses of neutral gauge bosons are shown as follows:
 \begin{align}
 \label{eq:G0}
 \begin{pmatrix}
	W^3_{R\mu}\\
	B'_{\mu}\\
	W^3_{L\mu}
\end{pmatrix} =&\left(
\begin{array}{ccc}
c_{\zeta } c_{\varphi }-s_W s_{\zeta } s_{\varphi } & -c_{\varphi } s_{\zeta }-c_{\zeta } s_W s_{\varphi } & c_W s_{\varphi } \\
-c_{\varphi } s_W s_{\zeta }-c_{\zeta } s_{\varphi } & s_{\zeta } s_{\varphi }-c_{\zeta } c_{\varphi } s_W & c_W c_{\varphi } \\
c_W s_{\zeta } & c_W c_{\zeta } & s_W \\
\end{array}
\right) \begin{pmatrix}
 	Z'_{\mu}\\
 	Z_{\mu}\\
 	A_{\mu}
 \end{pmatrix}, 
 \end{align}
where $t_{\varphi} \equiv g_{BL}/g_2$ , $s_{\varphi}= t_W$ is the condition matching to the SM gauge couplings to guarantee the  massless photon,  and 
\begin{align}
t_{2\zeta} =&\frac{2 c_{\varphi }^3 c_W v^2}{c_{\varphi }^2 v^2 \left(c_{\varphi }^2 c_W^2-1\right)+c_W^2 v_R^2}=  \frac{2 \left(2 c_W^2-1\right)^{3/2} v^2}{c_W^4 v_R^2+\left(4 c_W^4-6 c_W^2+2\right) v^2},
\crn m_Z^2 \simeq  & \frac{m_W^2}{c_W^2},\;
 m_{Z'}^2 \simeq \frac{m_{W'}^2}{c_{\varphi}^2}. 
\end{align}

 The Higgs potential is \cite{Ezzat:2021bzs}:
 \begin{align}
V_h= & \mu_1^2 \mathrm{Tr}(\phi^{\dagger} \phi) +  \mu_2^2 \left[\mathrm{Tr}( \phi^{\dagger} \tilde{\phi}) + \mathrm{Tr}(\phi \tilde{\phi}^{\dagger} ) \right] + \lambda_1 \left[  \mathrm{Tr}(\phi^{\dagger} \phi) \right]^2 + \lambda_2 \left[ \left( \mathrm{Tr}( \phi^{\dagger} \tilde{\phi}) \right)^2  + \left( \mathrm{Tr}(\phi \tilde{\phi}^{\dagger} )\right)^2  \right]
\crn &+ \lambda_3 \mathrm{Tr}( \phi^{\dagger} \tilde{\phi}) \mathrm{Tr}(\phi \tilde{\phi}^{\dagger} )  +\lambda_4  \mathrm{Tr}(\phi^{\dagger} \phi)  \left[ \mathrm{Tr}( \phi^{\dagger} \tilde{\phi}) + \mathrm{Tr}(\phi \tilde{\phi}^{\dagger} ) \right] + \mu_3^2 \left( \chi_R^\dagger \chi_R\right) + \lambda_5 \left( \chi_R^\dagger \chi_R\right)^2 
\crn &+ \alpha_1 \mathrm{Tr}(\phi^{\dagger} \phi)  \left( \chi_R^\dagger \chi_R\right) + \alpha_2 (\chi_R^\dagger \phi^{\dagger} \phi\chi_R) +\alpha_3 (\chi_R^\dagger \tilde{\phi}^{\dagger} \tilde{\phi} \chi_R) + \alpha_4 \left[ (\chi_R^\dagger \phi^{\dagger} \tilde{\phi} \chi_R) +\mathrm{H.c.}\right]. 
 \end{align}
The physical states corresponding masses and mixing  parameters of the model were shown previously \cite{Ezzat:2021bzs}; therefore, we do not repeat this in this work. The main results and notations used in this work are summarized in appendix \ref{app:LRIS}. The SM-like Higgs boson was also indicated to be consistent with  the experimental results.  
\subsection{ Couplings and  Feynman rules for LFV decays in LRIS}

The above ingredients lead  to the LFV couplings as follows:
\begin{align}
\mathcal{L}^{\ell \ell H} =& \left[  -  \frac{h}{2v} \overline{n_i} \left[ \lambda_{ij} P_L + \lambda^*_{ij} P_R \right]  n_j  -  \left(1+  \frac{h}{v} \right) \overline{e_R}\mathcal{M}^\ell e_L +\mathrm{ h.c.}\right]  
\crn& - \sum_{i=1}^9 \overline{n_{i}} \left\{ \left[ U^{\nu*}_{ai} \left( y^{L\dagger}s_{\beta} - \tilde{y}^{L\dagger}c_{\beta}\right)_{ab}  c_{\xi} +  U^{\nu*}_{(a+6)i}y^{s\dagger}_{ab} s_{\xi}  \right]  P_R 
\right. \crn& \left. \frac{}{}\hspace{1.6cm} + U^{\nu}_{(a+3)i} \left( y^Lc_{\beta} -\tilde{y}^L s_{\beta} \right)_{ab} c_{\xi} P_L 
\right\} e_{b} H^+  +\mathrm{ h.c.}+\dots,  \label{eq:LllH}
\\ \mathcal{L}^{\ell \ell V}= & -eA_{\mu} \overline{e_a}\gamma^{\mu}e_a 
\crn &+ \left[ \frac{g_2}{\sqrt{2}} W^{+\mu}   \sum_{i=1}^9\overline{n_{i}} \gamma^{\mu}\left( c_{\theta}U^{\nu*}_{ai}  P_L + s_{\theta}U^{\nu}_{(a+3)i} P_R \right)e_{a} +\mathrm{h.c.} \right] 
\crn &+ \left[ \frac{g_2}{\sqrt{2}} W'^{+\mu} \sum_{i=1}^9 \overline{n_{i}} \gamma^{\mu} \left( - s_{\theta} U^{\nu*}_{ai}  P_L + c_{\theta}U^{\nu}_{(a+3)i}  P_R \right)  e_{a} +\mathrm{h.c.} \right] 
\crn&+e Z^{\mu} \overline{e_a} \left[\left(c_{\zeta } t^{\mathrm{SM}}_L -\frac{s_{\zeta } s_W}{2 c_{\varphi } c_W^2} \right) P_L +\left( c_{\zeta } t^{\mathrm{SM}}_R + \frac{s_{\zeta } (c_{\varphi }^2c_W^2 -s_W^2)}{2 s_W c_{\varphi } c_W^2}\right) P_R \right] e_a
\crn&+\frac{e}{2} Z^{\mu}\sum_{i,j=1}^2 \overline{n_{i}}\left[ g^L_{Zij} P_L - g^{L}_{Zji} P_R\right]   n_j +\dots,  \label{eq:LllZ}
\end{align} 
where we have used $U^{\ell}_{L,R}=I_3$, 
\begin{align}
	\lambda_{ij}&=\lambda_{ji}=\sum_{c=1}^3\left(m_{n_i}U^{\nu*}_{ci} U^{\nu}_{cj} + m_{n_j}U^{\nu*}_{cj} U^{\nu}_{ci} \right) \to g^{R*}_{hij}=g^{L}_{hij}=\frac{g\delta_{hee}}{2m_W} \lambda_{ij}, 
\crn g^L_{Zij}&=\sum_{c=1}^3 \left[U^{\nu*}_{ci} U^{\nu}_{cj} \left( \frac{c_{\zeta }}{2 c_W s_W}  -\frac{s_{\zeta } s_W}{2 c_{\varphi } c_W^2}\right) +U^{\nu*}_{(c+3)i} U^{\nu}_{(c+3)j}  \frac{s_{\zeta } \left(c_{\varphi }^2 c_W^2+s_W^2\right)}{2 c_{\varphi } c_W^2 s_W} \right], \label{eq:gLZij}
\end{align}
and $t_{\xi}=vs_{2\beta}/v_R$ relating to the singly charge Higgs mixing defined in appendix \ref{app:LRIS}. 

It is  seen easily that the SM-like Higgs couplings $h\overline{e_a}e_b$ are LFV conservative, therefore LFV$h$ decays are loop-induced. In the numerical investigation, we focus on the case of $\mathcal{M}^{\ell}= \hat{\mathcal{M}}^{\ell}$ for simplicity. In addition, the $h \overline{n}n$  is the same as those in the previous simple ISS extension of the SM.  The same conclusion holds for the $Z\overline{n}n$ coupling  in the limit $s_{\zeta}=0$.   In the LRIS model, the particular couplings corresponding to Lagrangian parts  given in Eqs. \eqref{eq_LFh} and  \eqref{eq_LFV}, are  $g^{L,R}_{aFS}=g^{L,R}_{aiH^+}$ and $g^{L,R}_{aFV}=g^{L,R}_{aiW}, g^{L,R}_{aiW'}$ where
\begin{align}
	\label{eq:LRIScoupling}
	g^{L}_{aiH^+}=&   \frac{\sqrt{2}c_{\xi}}{v c_{2\beta}} \sum_{c=1}^3\left[ U^{\nu}_{(c+3)i}\left( \hat{\mathcal{M}^{\ell}} - m_D s_{2\beta} \right)_{ca} \right] ,
\crn 	g^{R}_{aiH^+}=&  \frac{\sqrt{2} c_{\xi}}{v c_{2\beta}} \sum_{c=1}^3 \left[ U^{\nu *}_{ci} \left( \hat{\mathcal{M}}^{\ell \dagger}s_{2\beta} - m_D^{\dagger}  \right)_{ca}  +    U^{\nu *}_{(c+6)i}(\hat{M}_R)_{ca} t^2_{\xi}\right],
	\crn	g^{L}_{aiW}=  & \frac{g_2}{\sqrt{2}} c_{\theta}U^{\nu*}_{ai}, \; 	g^{R}_{aiW}=  \frac{g_2}{\sqrt{2}} s_{\theta}U^{\nu}_{(a+3) i}, 
	\crn g^{L}_{aiW'}=  & -\frac{g_2}{\sqrt{2}} s_{\theta}U^{\nu*}_{ai}, \; 	g^{R}_{aiW'}=  \frac{g_2}{\sqrt{2}} c_{\theta}U^{\nu}_{(a +3)i},
\end{align}
where formulas of $y^L$ and $\tilde{y}^L$ given in Eq. \eqref{eq:yL} were used. In the numerical investigation, we will consider the simple case of  $m_D$ and $R_0$ with $U_N=I_3=\xi$ and the diagonal $\hat{\mu}_s$, which is enough to guarantee the $(g-2)_\mu$ data. The form factors relating to the one-loop contributions to $\Delta a_{e_a}$ and Br$(e_b  \to e_a \gamma)$ predicted by the LRIS model are shown in appendices \ref{subsec_g2} and \ref{subsec:LFV}.  The main one-loop contributions to $(g-2)_{e_a}$ anomalies and LFV decay rates predicted by the LRIS model originate from  the couplings given in Eq. \eqref{eq:LRIScoupling}, namely 
\begin{align}
\label{eq:cabRLRIS}
c^{\mathrm{LRIS}}_{(ab)R}& = c_{(ab)R}(H^+) +c_{(ab)R}(W) + c_{(ab)R}(W').
\end{align}   
The respective one-loop contributions to $a_{e_a}$ and Br$(e_b\to e_a \gamma)$ originating from the LRIS model are:
\begin{align}
	\label{eq:cLFVLRIS}
	a^{\mathrm{LRIS}}_{e_a} =& -\frac{4m_a}{e} \mathrm{Re} \left[c^{\mathrm{LRIS}}_{(aa)R}\right] -a^{\mathrm{SM}}_{e_a}(W),  
\crn	\mathrm{Br}(e_b\to e_a \gamma)^{\mathrm{LRIS}}=& \frac{48 \pi^2}{G_F^2m_b^2} \left(|c^{\mathrm{LRIS}}_{(ab)R}|^2 +|c^{\mathrm{LRIS}}_{(ba)R}|^2\right) \mathrm{Br}(e_b\to e_a \overline{\nu_a} \nu_b),
\end{align}
where $a^{\mathrm{SM}}_{e_a}(W)$ is the one-loop contribution from $W$ exchange predicted by the SM. 

Feynman rules for couplings of the SM-like Higgs boson with bosons relating  to LFV$h$ decays are shown in Table \ref{t:hXX}. 
\begin{table}[ht]
	\centering 
	\begin{tabular}{|c|c|c|c|}
		\hline 
		Vertex	& Coupling: & Vertex&Coupling\\ 
		\hline 
		$g_{h W^+W^-}$	&$g m_W(1-s_{2\beta } s_{2\theta }) \delta_{hee}^{-1}$& 	$g_{hW'^{+}W'^{-}}$ & $g m_W (s_{2\beta } s_{2\theta }+1) \delta_{hee}^{-1}$\\
		\hline 
		$g_{hW^{+}W'^{-}}$ & $g m_W s_{2\beta } \left(s_{\theta }^2-c_{\theta }^2\right) \delta_{hee}^{-1}$ &	&  \\
		\hline 
		$g_{ W^+H^-h}$ & $\frac{g}{2}  c_{\xi } s_{\theta } \left(c_{\beta }^2-s_{\beta }^2\right)$ &  $g_{W'^+H^- h}$ &  $\frac{g}{2}  c_{\theta } c_{\xi } \left(c_{\beta }^2-s_{\beta}^2\right)$	\\
		\hline
	\end{tabular}
	\caption{Vertex factors  for SM-like Higgs  couplings to charged Higgs and  gauge bosons in the  LRIS model.} \label{t:hXX}
\end{table}
The triple coupling $\lambda_{hH^+H^-}$ of the SM-like Higgs boson derived from the Higgs potential is:
\begin{align*}
\lambda_{hH^+H^-}=v \left[ \left(-2 s_{2\beta}^2  (2 \lambda_2+\lambda_3) +2 \lambda_1 \right) s_{\xi}^2 + c_{\xi}^2 \left(\alpha_1+\alpha_3+\alpha_4 s_{2\beta}+(\alpha_2-\alpha_3) \left(s_{2\beta}+s_{\beta}^2\right) \frac{}{}\right)\right]. 
\end{align*}
 Feynman rules for the  couplings of the gauge boson $Z$ to charged Higgs and gauge bosons associated with  one-loop contributions to the decay amplitude $Z\to e_b^+e_a^- $ are collected in  table \ref{table:Zxx}. 
\begin{table}[ht]
	\centering 
	\begin{tabular}{|c|c|}
		\hline
		Vertex& Coupling\\
		\hline
		$g_{ZH^+H^-}$	&$-\frac{ c_{\zeta} \left(s_{\xi}^2+2 s_W^2-1\right)}{2 s_Wc_W} +\frac{ s_{\zeta} \left(\left(s_{\xi}^2+2\right) s_W^2-1\right)}{2 s_Wc_W \sqrt{1-2 s_W^2}}$\\
				\hline 	
		$g_{H^-W^{+}Z}$& $\frac{ c_{2\beta} c_{\xi} m_W s_{\theta}}{s_W c_W}  \left(c_{\zeta}-\frac{s_{\zeta} s_W^2}{\sqrt{1-2 s_W^2}}\right)$ \\
		\hline 
		$g_{H^-W'^{+}Z}$& $\frac{ c_{2\beta} c_{\theta} c_{\xi} m_W}{s_W c_W} \left(c_{\zeta}-\frac{s_{\zeta} s_W^2}{\sqrt{1-2 s_W^2}}\right)$ \\
		\hline 
		$g_{ZW^{+}W^-}$& $ \frac{c_{\zeta} c_{\theta}^2}{t_W} -s_{\theta}^2 \left(c_{\zeta}  t_W+s_{\zeta} \sqrt{1-t_W^2}/s_W\right)$\\
		\hline 
			$g_{ZW'^{+}W'^-}$&$ c_{\zeta}  s_{\theta}^2t^{-1}_W -c_{\theta}^2 \left(c_{\zeta}  t_W+s_{\zeta} \sqrt{1-t_W^2}/s_W\right)$ \\
		\hline 
				$g_{ZW'^{+}W^-}, g_{ZW^{+}W'^-}$& $-  c_{\theta} s_{\theta} \left(\frac{c_{\zeta}}{s_Wc_W} + s_{\zeta} \sqrt{1-t_W^2}/s_W\right)$\\
		\hline 
		\end{tabular}
	\caption{Feynman rules for couplings of $Z$ to charged Higgs and gauge bosons.}\label{table:Zxx}
\end{table}

The above results for couplings of the SM-like Higgs $h$  and $Z$ show that  our results are consistent with the SM results in the limit $\theta,\zeta,\xi \varpropto v/v_R\to 0$, corresponding to the condition that $v_R\gg v$. Consequently, a number of couplings are suppressed, namely,  $g_{W^+H^-h}$,  $g_{H^-W^+Z}$, $g_{ZW'^+W^-}=g_{ZW^+W'^-} \to 0$, leading to the  weak constraint by experiments searching for decays $H^\pm \to W^\pm h$, $H^\pm \to W^\pm Z $, and $W'^\pm \to W^\pm Z $. On the other hand, the nonzero values of these mixing parameters  result in nonzero factors of  divergent parts in one-loop contributions to the LFV$h$ and  LFV$Z$ decay amplitudes. Therefore,  these one-loop contributions must be included to guarantee the divergent cancellation requirements in the total decay amplitudes, even if the finite parts may be tiny.   In numerical estimation, the one-loop contributions from diagrams containing very heavy gauge boson exchanges such as $W'$ are ignored, even when the couplings $hW'^+W'^-$, $hW^\pm W'^\mp$, and  $ZW'^+W'^-$ are orders of magnitude of the SM couplings even when $\theta =\zeta =\xi=0$. However, we can easily see  that the divergent parts are generally nonzero, therefore these one-loop contributions are still useful for checking the overall divergent cancellation in the final results.  

 \subsection{Numerical discussion}
 The numerical values of experimental data were taken from Ref. \cite{Workman:2022ynf}, including the neutrino oscillation data, masses of charged leptons; masses of two gauge bosons $W$,  $Z$, and the SM-like Higgs bosons, namely 
 \begin{align}
 	\label{eq_ex}
 	g &=0.652,\; G_F=1.166\times 10^{-5} \;\mathrm{GeV}^{-2},\;\alpha_e=\frac{1}{137}= \frac{e^2}{4\pi} ,\; s^2_{W}=0.231, 
 	\crn  m_W& =80.377 \; \mathrm{GeV},\;  m_Z =91.1876 \; \mathrm{GeV},\; m_h=125.25\; \mathrm{GeV}, \; 	 \Gamma_Z=2.4955 \; \mathrm{GeV}, 
 	\crn 	m_e&=5\times 10^{-4} \;\mathrm{GeV},\; m_{\mu}=0.105 \;\mathrm{GeV} ,\; m_{\tau}=1.776 \;\mathrm{GeV}. 
 \end{align}
 
 We will focus on the best-fit values  of the neutrino oscillation data~\cite{Workman:2022ynf} corresponding to  the normal order (NO) scheme  with $m_{n_1}<m_{n_2}<m_{n_3}$, namely 
 \begin{align}
 	\label{eq_d2mijNO}
 	&s^2_{12}=0.32,\;   s^2_{23}= 0.547,\; s^2_{13}= 0.0216 ,\; 
 	\crn &\Delta m^2_{21}\equiv m^2_{n_2} - m^2_{n_1}=7.55\times 10^{-5} [\mathrm{eV}^2], \;
 \crn&	\Delta m^2_{32} \equiv m^2_{n_3} - m^2_{n_2}=2.424\times 10^{-3} [\mathrm{eV}^2].
 \end{align}
On the other hand, we fix $ \delta= 180 \;[\mathrm{Deg}] $ for simplicity in numerical investigation. Consequently,   the neutrino masses and mixing matrix are fixed as follows:  
 \begin{align}
 	\hat{m}_{\nu}&=  \left( \hat{m}^2_{\nu}\right)^{1/2}= \mathrm{diag} \left( m_{n_1}, \; \sqrt{ m_{n_1}^2 + \Delta m^2_{21}},\; \sqrt{m_{n_1}^2  +\Delta m^2_{21} +\Delta m^2_{32}} \right), 	\label{eq:NOmnu}
 	\\  U_{\mathrm{PMNS}} &=f(\theta_{12},\theta_{13},\theta_{23},\delta) =\left(
 	\begin{array}{ccc}
 		c_{12} c_{13} & c_{13} s_{12} & s_{13} e^{-i \delta } \\
 		-c_{23} s_{12}-c_{12} s_{13} s_{23} e^{i \delta } & c_{13} c_{23}-s_{12} s_{13} s_{23} e^{i \delta } & c_{13} s_{23} \\
 		s_{12} s_{23}-c_{12} c_{23} s_{13} e^{i \delta } & -c_{23} s_{12} e^{i \delta } s_{13}-c_{13} s_{23} & c_{13} c_{23} \\
 	\end{array}
 	\right), 	\label{eq:NOupmns}
 \end{align}
 where $m_{n_1} \leq 0.035$ eV in order to guarantee the data of Plank 2018 \cite{Planck:2018vyg}.  
 
 Experimental data for $(g-2)_{e,\mu}$ anomalies   have been updated from Ref. \cite{Muong-2:2023cdq} showing a 5.1$\sigma$ standard deviation from the  SM prediction 
 %
  \cite{ Aoyama:2020ynm,Davier:2010nc, Danilkin:2016hnh,  Davier:2017zfy, Keshavarzi:2018mgv, Colangelo:2018mtw, Hoferichter:2019mqg, Davier:2019can, Keshavarzi:2019abf, Kurz:2014wya, Melnikov:2003xd, Masjuan:2017tvw, Colangelo:2017fiz, Hoferichter:2018kwz, Gerardin:2019vio, Bijnens:2019ghy, Colangelo:2019uex, Colangelo:2014qya, Blum:2019ugy, Aoyama:2012wk, Aoyama:2019ryr, Czarnecki:2002nt, Gnendiger:2013pva, Pauk:2014rta, Jegerlehner:2017gek, Knecht:2018sci,Eichmann:2019bqf, Roig:2019reh} that:  $\Delta a^{\mathrm{NP}}_{\mu}\equiv  a^{\mathrm{exp}}_{\mu} -a^{\mathrm{SM}}_{\mu} =\left(2.49\pm 0.48 \right) \times 10^{-9}$ \cite{Venanzoni:2023mbe}.  The  experimental $a_e$ data  was  reported from different groups~\cite{Hanneke:2008tm, Parker:2018vye, Morel:2020dww, Fan:2022eto}, predict the  same order of $|\Delta a^{\mathrm{NP}}_{e}|=\mathcal{O}(10^{-13})$ defined as the  deviation  between experiments and the SM prediction \cite{Aoyama:2012wj,  Laporta:2017okg, Aoyama:2017uqe,  Terazawa:2018pdc, Volkov:2019phy, Gerardin:2020gpp}. 
 
 The cLFV rates are constrained from recent experiments as follows  \cite{BaBar:2009hkt, MEG:2016leq, Belle:2021ysv, MEGII:2023ltw} :  	$\mathrm{Br}(\mu\rightarrow e\gamma) < 3.1\times 10^{-13}$, $\mathrm{Br}(\tau\rightarrow e\gamma) <3.3\times 10^{-8}$, and  $\mathrm{Br}(\tau\rightarrow \mu\gamma) <4.2\times 10^{-8}$.   The latest experimental constraints for LFV$h$ decay rates are $	\mathrm{Br}(h\rightarrow \tau \mu) <1.5\times 10^{-3}$, $	\mathrm{Br}(h\rightarrow \tau e) <2\times 10^{-3}$, and  $\mathrm{Br}(h\rightarrow \mu e) <4.4\times 10^{-5}$ \cite{CMS:2021rsq, ATLAS:2019xlq, CMS:2023pte, ATLAS:2023mvd}.  
 The latest experimental constraints for LFV$Z$ decay rates  are $	\mathrm{Br}(Z\rightarrow \tau^\pm \mu^\mp) <6.5\times 10^{-6} $,  $\mathrm{Br}(Z\rightarrow \tau^\pm e^\mp) <5.0\times 10^{-6} $, and $	\mathrm{Br}(Z\rightarrow \mu^\pm e^\mp) <2.62\times 10^{-7}$ \cite{ATLAS:2021bdj, ATLAS:2022uhq}.    In the following numerical investigation, we emphasize that all allowed points we collect for illustrations  simultaneously satisfy  the $1\sigma$ experimental range of $(g-2)_{\mu}$ data, $ 2.01 \times 10^{-9} \leq 	\Delta a^{\mathrm{LRIS}}_{\mu} \equiv \Delta a^{\mathrm{NP}}_{\mu} \leq 2.97\times 10^{-9}$, and all recent experimental  constraints of cLFV, LFV$h$ and LFV$Z$ decays mentioned above. The maximal value of $\Delta a^{\mathrm{LRIS}}_{e}\leq \mathcal{O}(10^{-14})$ predicted by our numerical result is smaller than that in recent experimental data.

 The unknown parameters of the LRIS model  will be scanned in the following ranges:
 \begin{align}
 \label{eq:scanranges}
  & v_R \in \left[ 10,\;100\right]\; [\mathrm{TeV}]; \; m_{H^\pm} \in \left[ 0.3,\; 5\right]\;  [\mathrm{TeV}]; \; M_{1,2,3} \in  \left[ 0.1,\; 10\right] \;  [\mathrm{TeV}];
 \crn 
  &m_{n_1} \in [10^{-3},0.035] \;  [\mathrm{eV}];\;  t_{\beta} \in [0.02,0.8];\;  x_0 \in[10^{-6},\; 5\times 10^{-4}];\;   \left( \hat{\mu}_s\right)_{11,33} \in [0.2,50],
 \end{align}
 and the matrix $\xi$ given in Eq. \eqref{eq:mD22} is  parameterized  as $\xi=-f(\xi_1,\xi_2,\xi_3,0)$ with scanning ranges  $|\xi_i|\leq \pi$.   The lower bound of $v_R$ is chosen based on the searches for the heavy gauge boson $W'$  at the  High Luminosity Large Hadron Collider (HL-LHC)  \cite{ThomasArun:2021rwf}. The upper bound of $x_0$ is constrained from the data of non-unitary of the active neutrino mixing matrix \cite{Fernandez-Martinez:2016lgt, Agostinho:2017wfs, Blennow:2023mqx}.

\begin{figure}[ht]
 	\centering
 	\begin{tabular}{ccc}
 		\includegraphics[width=5.5cm]{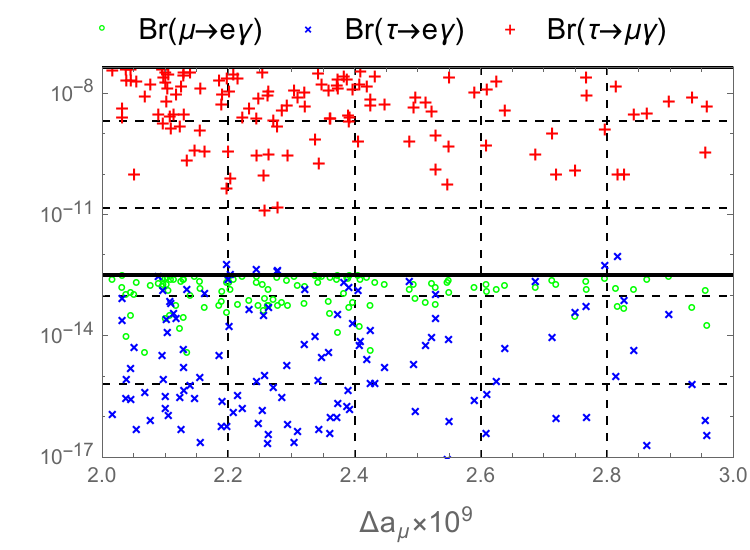}
 		&
 		\includegraphics[width=5.5cm]{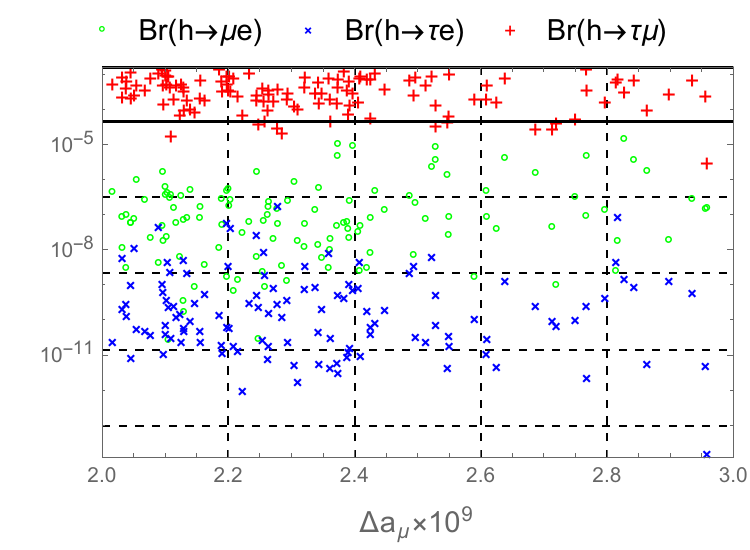} 
 		&
 		\includegraphics[width=5.5cm]{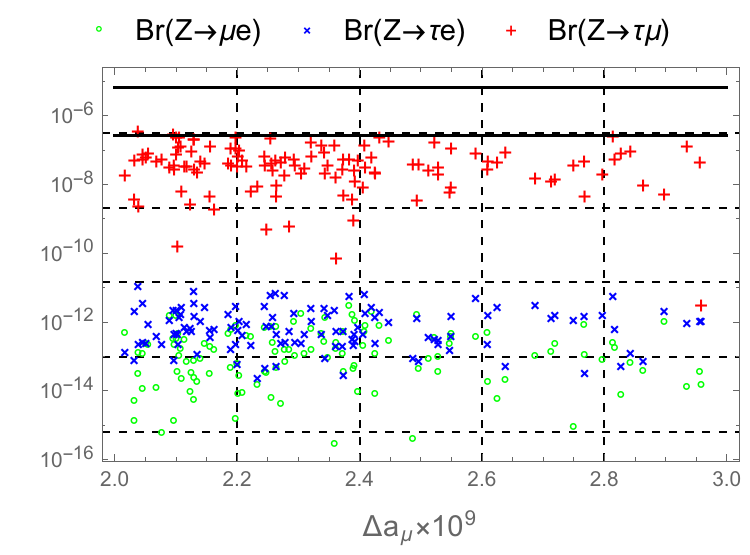}\\ 
 	\end{tabular}
 	\caption{The dependence of  LFV decay rates on $\Delta a_\mu$. Two black lines in the left, second, and right-panels show the upper bounds of LFV decay rates  for cLFV ($\tau \to \mu\gamma$, $\mu \to e\gamma$), LFV$h$ ($h\to \tau \mu,\mu e$), and LFV$Z$ ($Z\to \tau \mu,\mu e$), respectively. }\label{fig_amLFV}
 \end{figure}

 The correlations of $\Delta a_{\mu}$ with all LFV decay rates are illustrated numerically in Fig. \ref{fig_amLFV}.
 There are three decay rates that reach the experimental bounds, namely  Br$(\mu \to e\gamma)$, and Br$(\tau \to \mu \gamma)$ and Br$(h\to\tau \mu)$. Fortunately, they are still allowed for future sensitivities of Br$(\mu \to e\gamma)\simeq 6\times 10^{-14}$  \cite{MEGII:2018kmf}, Br$(\tau \to \mu\gamma)\simeq \mathcal{O}(10^{-9})$ \cite{Belle-II:2018jsg}, and  Br$(h\to\tau \mu) \simeq \mathcal{O}(10^{-4}) $ \cite{Qin:2017aju, Barman:2022iwj, Aoki:2023wfb, Jueid:2023fgo}. We conclude that the allowed regions of parameters of the LRIS model will be changed strongly by the incoming experimental results of the three mentioned LFV decays. 
 
 The upper bounds of the remaining LFV decay rates predicted by the LRIS model are:
 \begin{align}
\label{eq:LFVbound}
&\mathrm{Br}(\tau \to e\gamma) \leq 9\times 10^{-13}, \;  \mathrm{Br}(h \to\tau e) \leq 1.77\times 10^{-7},\;  \mathrm{Br}(h \to \mu e) \leq 1.3\times 10^{-5},
\crn& \mathrm{Br}(Z \to\mu^+ e^-) \leq 2.8\times 10^{-12},\;   \mathrm{Br}(Z \to\tau^+ e^-) \leq 1.1\times 10^{-11},\;  \mathrm{Br}(Z \to\tau^+ \mu^-) \leq 3.8\times 10^{-7}. 
 \end{align}
 Therefore, two decays $h\to \mu e$ and $Z\to \mu^+ \tau^-$ are close to incoming sensitivities \cite{Dam:2018rfz, FCC:2018byv}. 
 
 The allowed regions of the parameter space obtained in our investigation have stricter constraints than the scanning ranges chosen in Eq. \eqref{eq:scanranges}. In particular, the singly charged Higgs mass has an upper bound $m_{H^\pm} <4.2$ TeV, but it is still heavier than the values of interest in the popular 2HDMs  \cite{Han:2018znu, Barman:2021xeq, Athron:2021iuf, Wang:2022yhm}.   A hierarchy of heavy neutrino masses appears, namely   $m_{n_5}>2.8$ TeV, whereas  the two remaining masses have small upper bounds such that $m_{n_4},m_{n_6}<$ 380 GeV. 
 
 We note here that one-loop contributions from heavy charged gauge bosons $W'$ are suppressed with a fixed value $C_{UV}=0$, as in LoopTools.  Therefore,  total calculations, at least the divergent parts of all one-loop diagrams, including those relating to $W'$ exchanges, must be  considered. 
 
 We confirm here a property indicated in Ref. \cite{Ashry:2022maw} that the  dominant one-loop contributions to $\Delta a^{\mathrm{LRIS}}_{\mu}$ come from charged Higgs exchanges with the appearance of the chiral enhancement term proportional to $c^{\mathrm{LRIS}}_{(aa)R} \varpropto g^{L*}_{aiH^+} g^R_{aiH^+}$ \cite{Crivellin:2018qmi}.  Consequently, the values of $|c^{\mathrm{LRIS}}_{(ab)R}|$ with $a\neq b$ are large too. The numerical investigation showed that $c^{\mathrm{LRIS}}_{(ab)R}$  requires  strong destructive correlations between the Higgs and gauge boson contributions to ensure that all the LFV decay rates mentioned in this work satisfy the experimental constraints. Therefore, if the SM result for  $(g-2)_{\mu}$ in Ref. \cite{Borsanyi:2020mff} is accepted, implying  smaller values of $\Delta a^{\mathrm{LRIS}}_{\mu}$ than those chosen for our numerical illustration, the cancellation requirements for  $c^{\mathrm{LRIS}}_{(ab)R}$  will be relaxed, but the qualitative conclusions for LFV decay rates are unchanged.
 
 There are indirect sensitivities from other LFV processes such as the LFV  decays  $\mu \to 3 e$, and $\mu-e$ conversion in nuclei, as discussed in Refs. \cite{Alonso:2012ji} for the seesaw models.  In the LRIS framework, the $\mu-e$ conversion in nuclei was shown to be invisible  with recent experimental sensitivities \cite{Ashry:2022maw}. A similar conclusion is derived for the decay $\mu \to 3e$, because they have the similar one-loop contributions from the diagrams with  $Z$ exchanges. Another LFV indirect sensitivity is the LFV Higgs  $\mu-e$  couplings, which give two-loop contributions from Barr-Zee diagrams \cite{Barr:1990vd} to the  decay amplitude $\mu \to e \gamma$ \cite{Goudelis:2011un, Chang:1993kw, Davidson:2016utf} and tree-level decays $e_b\to 3e_a$. As we indicated in Eq. \eqref{eq:LllH}, the second term  implies that the two matrices expressing the SM-like Higgs couplings and charged lepton masses are proportional to each other. Consequently, the tree-level LFV  couplings of the SM-like Higgs boson do not appear in  the LRIS model under consideration, implying the absence of  two-loop Barr-Zee and tree-level diagrams with this Higgs exchange. In addition, the similar contributions from  heavy neutral Higgs exchange do not qualitatively change   the numerical results presented in this work.

\section{Conclusions}
\label{sec_concl}
In this work, we  completely introduce two classes of general master formulas expressing one-loop contributions to the LFV$h$ and LFV$Z$ decay amplitudes in the BSMs. The calculations were performed in the unitary gauge, independent of the couplings of nonphysical states such as Goldstone bosons. Analytical formulas are expressed in terms of the PV functions consistent with the notations introduced by LoopTools. The results show that the formulas corresponding to private one-loop diagrams generally contain divergent parts, which must vanish in the final finite amplitudes. Therefore, all diagrams containing divergences must be considered before ignoring them when their finite parts are estimated to be small with fixed divergence parts. The results introduced in this work are sufficient to estimate these divergences thoroughly.  We  numerically investigated the LFV$h$ and LFV$Z$ decay rates in the LSIS model, which accommodates all the data of neutrino oscillation, the cLFV decays, and the $(g-2)_{\mu}$ anomaly.  The results show that some of these decays are promising signals for incoming experimental searches. More importantly, the allowed regions of the parameter space are strongly affected by the recent experimental data from searches for three LFV decays  $\mu \to e\gamma$, $\tau  \to \mu\gamma$, and $h\to \tau \mu$. This implies that the allowed regions predicted by the LRIS model  change strongly once new LFV upper bounds are established.

\section*{ACKNOWLEDGMENT}
We thank the anonymous referee for pointing out an incorrect citation in the old manuscript. The referee's interesting comment on the indirect sensitivities of LFV four-body processes and two-loop contributions from Barr-Zee diagrams to the LFV decays will be carefully studied elsewhere. This research is funded by Vietnam National Foundation for Science and
Technology Development (NAFOSTED) under the grant number 103.01-2023.16.
\appendix
\section{Notations for Passarino-Veltman functions}
\subsection{General notations}
The PV-functions \cite{tHooft:1972tcz} used here to compute all form factors that give one-loop contributions  to the  LFV$h$ and LFV$Z$ decay amplitudes were listed in ref.~\cite{Hue:2017lak}, namely
\begin{align}
	&A_{0}(m^2)=\frac{(2\pi\mu)^{4-d}}{i\pi^2}\int \frac{d^d k}{k^2 -m^2 +i\delta},\crn
	&B_{\{0,\mu \}}(p^2_i,M^2_0,M^2_i) =\frac{(2\pi\mu)^{4-d}}{i\pi^2}\int \frac{d^d
		k \times\{1, k_{\mu}\}}{D_0 D_i}, \; i=1,2, \crn
	&C_{0,\mu,\mu\nu} \frac{(2\pi\mu)^{4-d}}{i\pi^2}\int \frac{d^d k \{ 1,k_\mu,k_\mu k_\nu\}}{D_0 D_1 D_2}, 
	\label{ABC_def}
\end{align} 
where  $D_0\equiv k^2-M_0^2  +i\delta$, $D_1\equiv (k -p_1)^2-M_1^2 + i\delta$, $D_2\equiv (k +p_2)^2 -M_2^2 + i\delta$, $C_{0,\mu,\mu\nu}=C_{0,\mu,\mu\nu}(p_1^2, q^2, p_2^2; M_0^2, M_1^2, M_2^2)$ with $q=p_1+p_2$,  and $\mu$ is an arbitrary mass parameter 
introduced via dimensional regularization \cite{tHooft:1972tcz}. The scalar PV-functions are defined consistent with LoopTools \cite{Hahn:1998yk}, namely:
\begin{align}
&A_0(m^2) =m^2(C_{UV} -\ln(m^2)+1),\; 
\crn	&B_{\mu}(p^2_i,M^2_0,M^2_1) =(-1)^ip_{1\mu}B_1(p^2_i,M^2_0,M^2_i),\; i=1,2;
	\crn &C_\mu = \left( -p_{1\mu}\right)C_1 +p_{2\mu} C_2,\crn
	&C_{\mu\nu} = g_{\mu\nu}C_{00} + p_{1\mu}p_{1\nu}C_{11} +p_{2\mu}p_{2\nu}C_{22}  -(p_{1\mu}p_{2\nu}+p_{2\mu}p_{1\nu})C_{12},
	\label{eq_fPVABC}
\end{align} 
where $C_{UV}=2/(4-d) -\gamma_E +\ln(4\pi \mu^2)$ are the divergent part. In our work, new  reduced notations will be used are $B^{(i)}_0\equiv B_0(p^2_i,M^2_0,M^2_i)$ and  $B^{(i)}_1 \equiv B_1(p^2_i,M^2_0,M^2_i)$.  
The scalar functions $A_0,B_0,C_0,C_{00},C_{i},C_{ij}$ ($i,j=1,2$) are well-known PV functions consistent with notations introduced in LoopTools \cite{Hahn:1998yk}. The scalar functions $A_0$, $B_0$, $C_0$ can be calculated using the techniques given in Ref. \cite{tHooft:1978jhc}.  Other PV functions needed in this work are 
\begin{align}
	&B_{0,\mu}(q^2;M_1,M_2)= \frac{(2\pi\mu)^{4-d}}{i\pi^2}\int \frac{d^d k~\{1,\; k_{\mu} \}}{D_1D_2},
	\crn&B_0^{(12)} \equiv B_0(q^2;M^2_1,M^2_2) =\dfrac{(2\pi\mu)^{4-d}}{i\pi^2}\int\dfrac{d^dk}{D_1D_2}= \dfrac{(2\pi\mu)^{4-d}}{i\pi^2}\int\dfrac{d^dk}{D'_1D'_2},
	\crn &B_\mu^{(12)} \equiv B_\mu(q^2; M^2_1,M^2_2) =\dfrac{(2\pi\mu)^{4-d}}{i\pi^2}\int\dfrac{d^dk\times k_\mu}{D_1D_2}\crn
	&\hspace{1.2cm}= \dfrac{(2\pi\mu)^{4-d}}{i\pi^2}\int\dfrac{d^dk\times (k +p_1)_{\mu}}{D'_1D'_2}= B_1^{(12)}q_\mu + B_0^{(12)}p_{1\mu}, 
\end{align}
where $B_1^{(12)}\equiv B_1(q^2; M_1^2,M_2^2)$, $ D'_1 \equiv k^{2}-M^2_1 +i\delta, D'_2 \equiv ((k+q)^2 -M^2_2 +i\delta $.

For simplicity, we define the following notations appearing in many important formulas:
\begin{align}
	\label{eq_Xidef}
	X_0&\equiv C_0 +C_1 +C_2,
	\crn   X_1&\equiv C_{11} +C_{12} +C_1
	\crn   X_2&\equiv C_{12} +C_{22} +C_2,
	\crn   X_3&\equiv C_1 +C_2=X_0-C_0,
	\crn   X_{012}&\equiv X_0+X_1 +X_2, \; X_{ij}= X_i+X_j.
\end{align}
The divergent parts of the PV-functions are:
\begin{align}
	\label{eq:divPV}
& \mathrm{div}[C_0] = \mathrm{div}[C_i] =\mathrm{div}[C_{ij}]=0; \; i,j=1,2, 
\crn&  \mathrm{div}[C_{00}]=\frac{C_{UV}}{4},  \;\mathrm{div}[B^{(1)}_{0}]=\mathrm{div}[B^{(2)}_{0}]= \mathrm{div}[B^{(12)}_{0}]=C_{UV} ,
\crn & \mathrm{div}[B^{(1)}_{1}]=\mathrm{div}[B^{(2)}_{1}]= \mathrm{div}[B^{(12)}_{1}]=- \frac{C_{UV}}{2}.
\end{align}
The Feynman rules for propagators of any gauge boson $V_{\mu}$ and their goldstone bosons in the  unitary gauge are as  follows 
\begin{align}
	\label{eq_DeltaVV}
	\Delta_{V}^{(u) \mu\nu}&	=\frac{-i}{k^2-m_V^2}\left( g^{\mu \nu} - \frac{k^{\mu} k^{\nu}}{ m^2_V}\right), \;
\Delta^{(u)}_{G_V} =0.
\end{align}
Before going to the details of the calculation. We list here important  well-known results such as the on-shell conditions gives $p_1^2=m_a^2$, $p_2^2=m_b^2$, and $q^2=m_Z^2$, where $m_a$, $m_b$, and $m_Z$ are the masses of leptons $a,b$ ($a,b=1,2,3$), and gauge boson $Z$. The momentum conservation gives $q=p_1 +p_2$. Two  internal momenta $k_1 \equiv k -p_1$ and  $k_2 \equiv k +p_2$ with $i=1,2$  are denoted in the diagram (1) of  Fig. \ref{fig_ZebaG}.

Then we take $d=4$ for all finite integrals.  For  all divergent integrals, after changing into the expressions in terms of the PV-functions, we take $d=4-2\epsilon$, then determining the final finite results before fixing $\epsilon=0$.  In addition, we will use the following transformation to change from integral to the notations of the PV-functions:
$$ \int{\dfrac{d^4k}{\left(2\pi\right)^4}} \to \frac{i}{16 \pi^2} \times \frac{(2\pi\mu)^{4-d}}{i\pi^2}\int d^dk= \frac{i}{16 \pi^2} \times \left( \mathrm{PV-functions} \right).$$
In practice, the overall factor $i/(16\pi^2)$ will be added in the final results.  Some intermediate steps  used in this work were presented precisely in Refs. \cite{Hue:2015fbb, Hong:2023rhg, Hong:2024yhk}.

\section{ \label{app_LFVU} General One-loop contributions to LFV$Z$ and LFV$h$ amplitudes in the unitary gauge}
\subsection{ \label{subsec_g2} Decays $e_b\to e_a \gamma$ and $(g-2)_{e_a}$} 
We use the analytic formulas for computing one-loop contributions to cLFV decay amplitudes and $(g-2)_{e_a}$  given in Ref. \cite{Crivellin:2018qmi}, which consistent with previous results \cite{Lavoura:2003xp, Hue:2017lak}. From the couplings given by two Eqs. \eqref{eq_LFh} and \eqref{eq_LFV},  the form factors $c_{(ab)R}$ corresponding to the one-loop contribution of a boson $X$ coupling with a fermion $F$ and a  charged lepton $e_a$ are:
\begin{align}
	\label{eq_cbaX}
	c^X_{(ab)R}&\equiv \frac{e}{16\pi^2 m^2_{X}} \left\{ g^{L*}_{aF X} g^{R}_{bF X} m_{F} \left[  f_X \left( t_X\right) +Q_F g_X \left( t_X\right)  \frac{}{}\right]  \right.  
	\crn  &  \left.  \hspace{2.2cm} +\left[\frac{}{}  m_{b} g^{L*}_{aF X} g^{L}_{bF X} + m_{a} g^{R*}_{aF X} g^{R}_{bF X} \right] \left[  \tilde{f}_X \left( t_X\right) +Q_F \tilde{g}_X \left( t_X\right) \right] \right\},  
\end{align}
where $X=S,V_{\mu}$, $t_X\equiv m^2_{F}/m^2_X$, $Q_F$ is the electric charge of the fermion $F$, and the master functions are 
\begin{align}
	\label{eq_MasterFunc}
	f_\Phi (x)&= 2\tilde{g}_\Phi(x)=\frac{x^2-1 -2x\ln x}{4(x-1)^3},\crn 
	g_\Phi&=\frac{x-1 -\ln x}{2(x-1)^2}, \crn 
	\tilde{f}_\Phi(x)&= \frac{2x^3 +3x^2 -6x +1 -6x^2 \ln x}{24(x-1)^4}, \crn
	f_V(x)&= \frac{x^3 -12 x^2 +15 x -4 +6 x^2\ln x}{4(x-1)^3}, \crn
	g_V(x)&=\frac{x^2-5x +4+3 x \ln x}{2(x-1)^2}, \crn
	\tilde{f}_V(x)&= \frac{-4x^4 +49x^3 -78 x^2 +43x -10 -18x^3\ln x}{24(x-1)^4}, \crn
	\tilde{g}_V(x)&= \frac{-3(x^3 -6x^2 +7x -2 +2x^2\ln x)}{(x-1)^3}. 
\end{align}

Formulas for one-loop contributions to $a_{e,\mu}$ and cLFV decay rates are:
\begin{align}
\label{eq:cLFV}
	a_{e_a}(X) =& -\frac{4m_a}{e} \mathrm{Re} \left[c^X_{(aa)R}\right],  \\
	\mathrm{Br}(e_b\to e_a \gamma) =& \frac{48 \pi^2}{G_F^2m_b^2} \left(|c_{(ab)R}|^2 +|c_{(ba)R}|^2\right) \mathrm{Br}(e_b\to e_a \overline{\nu_a} \nu_b),
\end{align}
where $G_F=g^2/(4\sqrt{2}m_W^2)$ is the Fermi constant, Br$(\mu \to e\overline{\nu_e} \nu_\mu) \simeq 1$, Br$(\tau \to e\overline{\nu_e} \nu_\tau) \simeq 0.1782$, Br$(\tau \to \mu\overline{\nu_\mu} \nu_\tau) \simeq0.1739$, and $c_{(ab)R}=\sum_X c_{(ab)R}(X)$ with $X$ being all relevant bosons predicted by the particular BSM.  

In the LRIS model, precise expressions of $c_{(ab)R}$ with $U_N=I_3$ are
\begin{align}
	\label{eq:cabR0}
	c_{(ab)R}(H^+) =& \frac{e}{16 \pi^2 m^2_{H^+}}\sum_{i=1}^9 \left[ g^{L*}_{aiH^+}g^{R}_{biH^+}m_{n_i} f_{\Phi} (x_{i,H}) + \left( m_b g^{L*}_{aiH^+}g^{L}_{biH^+} +m_a g^{R*}_{aiH^+}g^{R}_{biH^+}\right) \tilde{f}_{\Phi}(x_{i,H})\right], 
	\crn	c_{(ab)R}(W^+) =&\frac{eg^2_2}{32 \pi^2 m^2_{W}}\sum_{i=1}^9 \left[ s_{\theta}c_{\theta} U^{\nu}_{ai} U^{\nu}_{(b+3)i}m_{n_i} f_{V} (x_{i,W}) 
	\right. \crn&\left. \hspace{2.4cm} 
	+ \left( m_b  c^2_{\theta}U^{\nu}_{ai} U^{\nu*}_{bi} +m_a s^2_{\theta}U^{\nu}_{(a+3)i} U^{\nu*}_{(b+3)i}\right) \tilde{f}_{V}(x_{i,W})\right], 
	\crn	c_{(ab)R}(W'^+) =& \frac{eg^2_2}{32 \pi^2 m^2_{W'}}\sum_{i=1}^9 \left[ -s_{\theta}c_{\theta} U^{\nu}_{ai} U^{\nu}_{(a+3)i}m_{n_i} f_{V} (x_{i,W'}) 
	\right. \crn&\left. \hspace{2.4cm} 
	+ \left( m_b  s^2_{\theta}U^{\nu}_{ai} U^{\nu*}_{ai} +m_a c^2_{\theta}U^{\nu}_{(a+3)i} U^{\nu*}_{(a+3)i}\right) \tilde{f}_{V}(x_{i,W'})\right], 
\end{align}
where $x_{i,B}\equiv m_{n_i}^2/m_B^2$ with $B=H^\pm,W,W'$.

\subsection{Decays $Z\to e_a^\pm e_b^\mp$} 

For simplicity,  we will use  new notations of products of two LFV couplings introducing in Eqs. \eqref{eq_LFh} and  \eqref{eq_LFV} that $g^{XY}_{FBB'} \equiv  g^{X*}_{aFB} g^{Y}_{bFB'}$ with $X,Y=L,R$ and $B,B'=V,V',S,S'$ being charged Higgs and gauge boson exchanges in  particular diagrams. The corresponding arguments for PV-functions are $(m_a^2,q^2,m_b^2; m_F^2,m_B^2,m_{B'}^2)$, which is used to identify with the LoopTools notations that  $C_{x}=C_x(m_a^2,q^2,m_b^2; m_F^2,m_B^2,m_{B'}^2)$ for $x=0,i,00, ij$ ($i,j=1,2$), $B^{(1)}_{0,1}=B_{0,1}(m_a^2;m_F^2,m_B^2)$, $B^{(2)}_{0,1}=B_{0,1}(m_b^2;m_F^2,m_{B'}^2)$, and $B^{(12)}_{0,1}=B_{0,1}(q^2;m_B^2,m_{B'}^2)$.   The second notation is  $g^{XY}_{BFF'} \equiv  g^{X*}_{aFB} g^{Y}_{bF'B}$ with $B=S,V$ corresponding the  argument  $(m_a^2,q^2,m_b^2; m_B^2, m_F^2, m_{F'}^2)$ for PV-functions used in LoopTools that  $C_{x}=C_x(m_a^2,q^2,m_b^2; m_B^2,m_F^2,m_{F'}^2)$,  $B^{(1)}_{0,1}=B_{0,1}(m_a^2;m_B^2, m_F^2)$, $B^{(2)}_{0,1}=B_{0,1}(m_b^2;m_{B}^2, m_{F'}^2)$, and $B^{(12)}_{0,1}=B_{0,1}(q^2;m_F^2,m_{F'}^2)$. In particular LFV$h$ or   LFV$Z$ decays with $q^2=m_h^2,m_Z^2$, we  will  pay attention to the last three parameters in the arguments under consideration $(m_F^2,m_B^2,m_{B'}^2)$ or $(m_B^2, m_F^2, m_{F'}^2)$. 

Regarding to the LFV$Z$ decay, form factors for one-loop contribution from diagram (1)    in Fig. \ref{fig_ZebaG} are: 
\begin{align}
	\bar{a}^{FVV'}_L =&  g_{ZVV'} \left\{ g^{LL}\left[ \left( 2(2-d) +m_F^2 f\right) C_{00} +2(m_Z^2 -m_a^2 -m_b^2)X_3
	\frac{}{}\right. \right. \crn& \left. \left.\frac{}{} \hspace{2cm}-\left( f (m_V^2 +m_{V'}^2)+4\right) (B^{(12)}_0 +m_F^2 C_0)  
	\right. \right. \crn& \left. \left.\frac{}{} \hspace{2cm} + \frac{1}{m_V^2} \left( A_0(m_{V}) +m_F^2 B^{(1)}_0 +m_a^2 B^{(1)}_1 -(m_V^2 - m_{V'}^2 +m_Z^2) m_a^2 C_1\right)  
	\right. \right. \crn& \left. \left.\frac{}{} \hspace{2cm} + \frac{1}{m_{V'}^2} \left( A_0(m_{V'}) +m_F^2 B^{(2)}_0 +m_b^2 B^{(2)}_1 -(-m_{V}^2 +m_{V'}^2 +m_Z^2) m_b^2 C_2 \right)\right]
	\right. \crn&\left. \qquad \quad +g^{RR} m_am_b\left[f \left(C_{00} +m_{V'}^2C_2  +m_{V}^2C_1\right)-2 X_3 \frac{}{}\right]	
	\right. \crn&\left. \qquad \quad - g^{RL} m_a m_F \left[f C_{00} +(2 -fm_{V'}^2)C_0 +f (m_V^2 -m_{V'}^2)C_1 +\frac{B^{(1)}_0 +B^{(1)}_1}{m_V^2}\right]	
	\right. \crn&\left. \qquad \quad - g^{LR}m_b m_F \left[f C_{00} +(2-f m_{V}^2)C_0 -f(m_V^2 -m_{V'}^2)C_2 +\frac{B^{(2)}_0 +B^{(2)}_1}{m_{V'}^2}\right]	
	\right\},  \label{eq:avvfL} 
	%
	\\ \bar{a}^{F VV'}_R  =& \bar{a}^{FVV'}_L\left[ g^{LL} \to g^{RR}, g^{RL} \to g^{LR},\; g^{LR} \to g^{RL}\right] , \label{eq:avvfR}
	\\ \bar{b}^{F VV'}_L  =& g_{ZVV'} \left\{ g^{LL} m_a\left[ 4(X_3 -X_1) +f (m_F^2 X_{01} +m_b^2 X_2) -2 (m_{V'}^2f +2) C_2\right]
	\frac{}{}\right. \crn&\left. \qquad \;\; + g^{RR}m_b \left[ 4(X_3 -X_2) +f (m_F^2 X_{02} +m_a^2 X_1) -2 (m_{V}^2f +2) C_1\right]	
	\right. \crn&\left. \qquad \;\;  - g^{RL}  m_F \left[ f \left( \frac{}{} m_F^2 X_0 +m_a^2 X_1 +m_b^2 X_2 -2 m_{V}^2 C_1 -2 m_{V'}^2 C_2\right) + 4 X_3 \right]	
	\right. \crn&\left. \frac{}{} \qquad  - g^{LR} m_a m_b m_F f X_{012} 	
	\right\},   \label{eq:bvvfL} 
	%
	\\ \bar{b}^{FVV'}_R =& \bar{b}^{FVV'}_L\left[ g^{LL} \to g^{RR}, g^{RR} \to g^{LL}, g^{RL} \to g^{LR},\; g^{LR} \to g^{RL}\right], \label{eq:bvvfR}
\end{align} 
where  $g^{XY}\equiv g^{XY}_{FVV'}$,  the arguments for PV-funtions are $( m_F^2, m_{V}^2 ,m_{V'}^2)$, and   
$$ f=  \frac{(m_Z^2 -m_V^2 -m^2_{V'})}{m_V^2 m_{V'}^2}.$$

The diagram (2) in Fig. \ref{fig_ZebaG}  corresponding to the following form factors:
\begin{align}
	\bar{a}^{VFF'}_L =& \frac{g^{LL}}{m_V^2} \left\lbrace g^L_{ZFF'} \left[  m^2_V \Big((2-d)^2C_{00}  +2 m^2_aX_{01} +2m^2_bX_{02} -2m^2_Z \left( C_{12}+X_0\right)\Big) \right.\right.\crn
	&\left.\left. \hspace{2.2cm} -A_0(m_V)- \left(m_{F}^2 -m^2_a\right)B_0^{(1)} -\left(m_{F'}^2-m^2_b\right)B_0^{(2)}  +m^2_aB_1^{(1)} +m^2_bB_1^{(2)}
	\right.\right.\crn
	&\left.\left. \hspace{2.2cm}-m_a^2m_b^2 X_0  -m_{F}^2 m_{F'}^2C_0  +m_{a}^2 m_{F}^2 (C_0 +C_1)	+m_{b}^2 m_{F'}^2 (C_0 +C_2)   \frac{}{}\right] 
	\right.\crn	&\left. \hspace{0.8cm}-g^R_{ZFF'} m_{F} m_{F'}\bigg[2m^2_V C_0  +(2-d)C_{00} -m^2_a(X_1 -C_{1}) -m^2_b(X_{2} -C_2) +m^2_Z C_{12}\bigg]\right\rbrace
	\crn &+ \frac{g^{RR}g^R_{ZFF'} m_am_b}{m_V^2}
	\crn& \times  \left[ \frac{}{}  2m_{V}^2X_0 -(2-d) C_{00} +m^2_aX_1 +m^2_bX_2 	 -m_Z^2 C_{12}  -m_{F}^2C_1 -m_{F'}^2C_2 \right],  \label{eq_al2u} 
	%
	\\ \bar{a}^{VFF'}_R  =& \bar{a}^{VFF'}_L \left[g^{LL}\to g^{RR}, g^{RR}\to g^{LL},  g^L_{ZFF'} \to  g^R_{ZFF'}, g^R_{ZFF'} \to  g^L_{ZFF'}\right], \label{eq_ar2u}
	\\ \bar{b}^{VFF'}_L  =& \frac{2g^{LL} m_a}{m_V^2}  \left[ g^L_{ZFF'} \left(-2m_{V}^2 X_{01}   -m^2_bX_2 +m_{F'}^2C_2\right) -g^R_{ZFF'} m_{F}m_{F'}(X_1 -C_1) \frac{}{}\right]
	\crn &+  \frac{2g^{RR} m_b}{m_V^2}  \left[ g^R_{ZFF'} \left(-2m_{V}^2 X_{02}   -m^2_aX_1 +m_{F}^2C_1\right) -g^L_{ZFF'} m_{F}m_{F'}(X_2 -C_2) \frac{}{}\right],   \label{eq_bl2u} 
	%
	\\ \bar{b}^{VFF'}_R =&  \bar{b}^{VFF'}_L \left[g^{LL}\to g^{RR}, g^{RR}\to g^{LL},  g^L_{ZFF'} \to  g^R_{ZFF'}, g^R_{ZFF'} \to  g^L_{ZFF'}\right], \label{eq_br2u}
\end{align} 
where  $g^{XY}\equiv g^{XY}_{VFF'}$ and arguments for PV-funtions are $( m_V^2,m_{F}^2, m_{F'}^2)$.

The sum of two diagrams (7) and (8) gives the following form factors:
\begin{align}
	\bar{a}_L^{FV} =&  \dfrac{ t_L}{(m_a^2 -m^2_b) m_V^2}
	%
	\left\lbrace  g^{LL} \left[ \left((d-2) m_V^2 +m_F^2\right) \left(m_a^2 B^{(1)}_1 -m_b^2 B^{(2)}_1 \right) + m_a^4 B^{(1)}_1 -m_b^4 B^{(2)}_1
	\right. \right. \crn &\left. \left. \hspace{3.7cm}+ (m_a^2 -m_b^2)A_0(m_V) + 2 m_F^2 \left(m_a^2 B^{(1)}_0 -m_b^2 B^{(2)}_0 \right)  \right] 
	\right.\crn&\left. \hspace{2.9cm} +g^{RR} m_am_b \left[ \left(2m_V^2 +m_F^2\right) \left( B^{(1)}_1 - B^{(2)}_1 \right) 
	\right. \right. \crn &\left. \left. \hspace{5.1cm}
	+ m_a^2 B^{(1)}_1 -m_b^2 B^{(2)}_1 + 2 m_F^2 \left( B^{(1)}_0 - B^{(2)}_0 \right)  \right] 
	\right.\crn&\left.  \hspace{2.9 cm} +3\left( m_a g^{RL} +m_b g^{LR}\right)m_F m_V^2  \left( B^{(1)}_0 -B^{(2)}_0 \right)
	\right\rbrace,
	\crn \bar{a}_R^{FV} =& \bar{a}_L^{FV} \left[ t_L\to t_R, g^{LL} \to g^{RR}, g^{RR} \to g^{LL}, g^{RL} \to g^{LR},\; g^{LR} \to g^{RL}\right],
	\crn \bar{b}_L^{FV} =& \bar{b}_R^{FV} =0, 
\end{align}
where  $g^{XY} \equiv g^{XY}_{FVV}$, and  $B^{(k)}_{0,1}=B_{0,1}(p_k^2;m_F^2,m_V^2)$.

One-loop form factors from diagram (3)    in Fig. \ref{fig_ZebaG} are:  
\begin{align}
	\bar{a}^{FVS}_L =& \frac{g^*_{SVZ}}{m_V^2} \left[ \frac{}{}g^{LL} m_F (m_V^2 C_0 -C_{00}) + g^{RL} m_a (m_V^2 C_1 +C_{00}) -g^{LR} m_bm_V^2 C_2\right],  \label{eq_al2u} 
	%
	\\ \bar{a}^{FVS}_R  =& \bar{a}^{FVS}_L\left[\frac{}{} g^{LL} \to g^{RR}, g^{RL} \to g^{LR},\; g^{LR} \to g^{RL}\right] , \label{eq_ar2u}
	\\ \bar{b}^{FVS}_L  =& \frac{g^*_{SVZ}}{m_V^2} \left[ -m_F \left(g^{LL}m_a X_{01} +g^{RR} m_b X_2\right) 
	\frac{}{}\right. \crn& \left. \frac{}{}\qquad + g^{RL} \left( -2 m_V^2 C_1 +m_a^2 X_1 +m_F^2 X_0\right) + g^{LR} m_am_b X_2 \right],   \label{eq_bl2u} 
	%
	\\ \bar{b}^{FVS}_R =& \bar{b}^{FVS}_L\left[\frac{}{} g^{LL} \to g^{RR}, g^{RR} \to g^{LL}, g^{RL} \to g^{LR},\; g^{LR} \to g^{RL}\right], \label{eq_br2u}
\end{align} 
where  $g^{XY} \equiv g^{XY}_{FVS}$ and arguments for PV-funtions are $( m_F^2,m_{V}^2, m_{S}^2)$.  

One-loop form factors from diagram (4)    in Fig. \ref{fig_ZebaG} are: 
\begin{align}
	\bar{a}^{FSV}_L =& \frac{g_{SVZ}}{m_V^2} \left[ g^{LL} m_F (m_V^2 C_0 -C_{00}) + g^{RL} m_b (m_V^2 C_2 +C_{00}) -g^{LR} m_am_V^2 C_1\right],  \label{eq_al2u} 
	%
	\\ \bar{a}^{FSV}_R  =& \bar{a}^{FSV}_L\left[ g^{LL} \to g^{RR}, g^{RL} \to g^{LR},\; g^{LR} \to g^{RL}\right] , \label{eq_ar2u}
	\\ \bar{b}^{FSV}_L  =& \frac{g_{SVZ}}{m_V^2} \left[ -m_F \left(g^{RR}m_b X_{02} +g^{LL} m_a X_1\right) 
	\right. \crn& \left. \frac{}{}\qquad + g^{LR} \left( -2 m_V^2 C_2 +m_b^2 X_2 +m_F^2 X_0\right) + g^{RL} m_am_b X_2 \right],   \label{eq_bl2u} 
	%
	\\ \bar{b}^{FSV}_R =& \bar{b}^{FVS}_L\left[ g^{LL} \to g^{RR}, g^{RR} \to g^{LL}, g^{RL} \to g^{LR},\; g^{LR} \to g^{RL}\right], \label{eq_br2u}
\end{align} 
where  $g^{XY} \equiv g^{XY}_{FSV}$ and arguments for PV-funtions are $( m_F^2, m_{S}^2 ,m_{V}^2)$.

The contributions from diagrams with pure scalar exchanges were shown previously in Ref. \cite{Jurciukonis:2021izn, Hong:2023rhg}.  Particular formulas of the amplitudes are written as follows.  Final results for form factors corresponding to diagram (5) are:
\begin{align}
	\label{eq_ab4LR}	
	\bar{a}^{FSS'}_{L}&=- 2g_{ZS'^*S} g^{LL} C_{00},
	\crn  \bar{a}^{FSS'}_{R}& = -2g_{ZS'^*S} g^{RR}C_{00},
	\crn  \bar{b}^{FSS'}_{L}& = -2g_{ZS'^*S} \left[   m_a  g^{LL} X_1  +  m_b  g^{RR} X_2  -m_{F} g^{RL}  X_0  \right] ,
	\crn  \bar{b}^{FSS'}_{R}& =-2g_{ZS'^*S} \left[  m_a  g^{RR}  X_1  +  m_b  g^{LL}X_2   -m_{F} g^{LR} X_0  \right],
\end{align}
where arguments for PV-funtions are $(m_{F}^2, m_{S}^2, m_{S'}^2)$, and $g^{XY} = g^{XY}_{FSS'}$.

Forms factors corresponding to diagram (6) are
\begin{align}
	\label{eq_ab6LR}	
	\bar{a}^{SFF'}_{L}=& - \left\{  g^{L}_{ZFF'}\left[ g^{LL}  m_{F} m_{F'}C_0 + g^{RL}  m_{a} m_{F'}(C_0+C_1) \frac{}{}
	\right.\right.\crn&\left.\left.\qquad \qquad \qquad
	+  g^{LR}  m_{b} m_{F}(C_0+C_2) +  g^{RR} m_{a} m_bX_0 \frac{}{}\right]   
	\right.\crn  &\left. \qquad \quad-g^{R}_{ZFF'} \left[ g^{LL} \left( (d-2)C_{00} +m_a^2 X_1 +m_b^2X_2 -m_Z^2 C_{12}\right) \frac{}{}
	\right.\right.\crn&\left.\left.\qquad \qquad \qquad \; \frac{}{} +m_am_{F}g^{RL}C_1 + m_bm_{F'}g^{LR}C_2 \right]\right\}
	\crn  \bar{a}^{SFF'}_{R}=& - \left\{  -g^{L}_{ZFF'} \left[ g^{RR} \left( (d-2) C_{00} +m_a^2 X_1 +m_b^2X_2 -m_Z^2 C_{12}\right) \frac{}{}
	\right.\right.\crn&\left.\left.\qquad \qquad \qquad\; +g^{LR} m_am_{F_1}C_1 +g^{RL}m_bm_{F_2}C_2 \frac{}{}\right]
	\right.\crn  &\left. \qquad \quad +
	\frac{}{} g^{R}_{ZFF'}\left[ g^{RR}  m_{F} m_{F'}C_0 + g^{LR}  m_{a} m_{F'}(C_0+C_1) \frac{}{}
	\right.\right.\crn&\left.\left.\qquad \qquad \qquad \frac{}{}+  g^{RL}  m_{b} m_{F}(C_0+C_2) +  g^{LL}  m_{a} m_bX_0\right] 	\right\},
	\crn \bar{b}^{SFF'}_{L}=& -2 \left[\frac{}{} g^{L}_{ZFF'}  \left( g^{RL} m_{F'}C_2 +  g^{RR}  m_{b} X_2\right) 
	%
	+g^{R}_{ZFF'}  \left(  g^{RL}  m_{F}C_1 +  g^{LL}  m_{a} X_1\right) 
	\right]	,
	\crn\bar{b}^{SFF'}_{R} =&- 2 \left[\frac{}{} g^{L}_{ZFF'}   \left(  g^{LR} m_{F}C_1 +  g^{R R}  m_{a} X_1\right)
	%
	+g^{R}_{ZFF'} \left( g^{LR} m_{F'}C_2 +  g^{LL}  m_{b} X_2\right) 
	\right], 
\end{align}
where $g^{XY} \equiv g^{XY}_{SFF'}$  and   arguments for PV-funtions are $(m_S^2,m^2_{F},m^2_{F'})$.  

 Sum of two diagrams (9) and (10) gives the following form factors 
\begin{align}
	\label{eq_a910LR}
	\bar{a}^{FS}_{L} = & - \frac{ t_{L}}{m_a^2 -m_b^2} \left[  m_{F} \left( m_a g^{RL}  + m_b g^{LR}   \right) \left(B^{(1)}_0 -B^{(2)}_0\right) 
	\right. \crn& \left. \hspace{2.2cm} -m_am_b g^{RR}\left( B^{(1)}_1-B^{(2)}_1\right)   - g^{LL}   \left(m_a^2B^{(1)}_1 -m_b^2 B^{(2)}_1 \right) 
	\right],
	\crn \bar{a}^{FS}_{R}&= 	\bar{a}^{FS}_{L} \left[ t_L\to t_R, g^{LL} \to g^{RR}, g^{RR} \to g^{LL}, g^{RL} \to g^{LR},\; g^{LR} \to g^{RL}\right],
\crn \bar{b}^{FS}_{L}&= \bar{a}^{FS}_{R}=0,
\end{align}
where $g^{XY}\equiv g^{XY}_{FSS} $  and  $B^{(k)}_{0,1}=B_{0,1}(p_k^2;m_F^2,m_S^2)$. 

\subsection{One-loop contributions to decays $h\to e_a^\pm e_b^\mp$} 

The diagram (1) in Fig. \ref{fig_hebaG}  corresponding to the following amplitude:
\begin{align}
	\Delta^{FVV'}_L=& -\frac{g_{hVV'} g^{LL} m_a}{16\pi^2} \left\{  2 C_1 -\frac{1}{m_V^2} \left[  B^{(12)}_1 -(m_F^2 -m_a^2) (C_0 +C_1)  -(B^{(2)}_0  +m_V^2 C_0)   \right] 
	\right.\crn&\left. \hspace{2.95cm} -\frac{1}{m_{V'}^2} \left[ B^{(12)}_1 +B^{(12)}_0 -(m_F^2 -m_b^2) C_1 \right] 
	\right.\crn&\left. \hspace{2.95cm} - \frac{1}{2 m_V^2 m_{V'}^2} \left[ A_0(m_{V'})   +m_F^2 \left( B^{(1)}_0 + B^{(2)}_0+B^{(1)}_1 \right) +  m_b^2 B^{(2)}_1
	\right. \right. \crn&\left. \left. \hspace{5.0cm}    
	+ (m_V^2 +m_{V'}^2 -q^2) \left(m_F^2(C_0 +C_1) + m_b^2 C_2- B^{(12)}_1\right) \right]  \right\}
	\crn &- \frac{g_{hVV'}  g^{RR} m_b}{16\pi^2} \left\{ 2C_2 +\frac{1}{m_V^2} \left[ B^{(12)}_1 +(m_F^2 -m_a^2) C_2 \right] 
	\right. \crn&\left. \hspace{2.85 cm}+\frac{1}{m_{V'}^2} \left[B^{(12)}_1 +B^{(12)}_0 +(m_F^2 -m_b^2) (C_0 +C_2) +B^{(1)}_0  +m_{V'}^2 C_0  \right] 
	\right. \crn&\left. \hspace{2.95cm}  - \frac{1}{2 m_V^2 m_{V'}^2} \left[  A_0(m_{V})    + m_F^2 \left( B^{(1)}_0+ B^{(2)}_0 +B^{(2)}_1 \right)
	+  m_a^2 B^{(1)}_1
	\right. \right.  \crn&\left. \left. \hspace{5cm} + (m_V^2 +m_{V'}^2 -q^2)   \left(m_F^2(C_0 +C_2) +m_a^2C_1+B^{(12)}_1 +B^{(12)}_0 \right)  \right] \right\}
	\crn &-\frac{g_{hVV'} g^{RL} m_F}{32\pi^2 m_V^2 m_{V'}^2} \left\{ 4 m_V^2 m_{V'}^2 C_0 +A_0(m_V) +A_0(m_{V'}) +(m_F^2 -2 m_V^2) B^{(1)}_0  
	\frac{}{}\right.\crn& \left. \frac{}{} \hspace{2.8cm} +(m_F^2 -2 m_{V'}^2) B^{(2)}_0  + m_a^2 B^{(1)}_1 + m_b^2 B^{(2)}_1 
	\frac{}{}\right.\crn& \left. \frac{}{} \hspace{2.8cm} +(m_V^2 +m_{V'}^2 -q^2) (B^{(12)}_0 +m_a^2 C_1 +m_b^2 C_2 +m_F^2 C_0)  \right\}
	\crn &-\frac{g_{hVV'}  g^{LR} m_am_bm_F}{32 \pi^2 m_V^2 m_{V'}^2}  \left[B^{(1)}_0 +B^{(2)}_0 + B^{(1)}_1 +B^{(2)}_1  + (m_V^2 +m_{V'}^2 -q^2) X_0\right], \label{eq:DeFVVL}
	\\\Delta^{FVV'}_R=&  \Delta^{FVV'}_L\left[ g^{LL} \to g^{RR},  g^{RR} \to g^{LL},  g^{RL} \to g^{LR},  g^{LR} \to g^{RL}\right],\label{eq:DeFVVR}
\end{align}
where the coupling notations are $g^{XY}=g^{XY}_{FVV'}$, $q^2=m_h^2$, and the arguments of PV-functions is $(m_F^2,m_V^2,m_{V'}^2)$.

The diagram (2) in Fig. \ref{fig_hebaG}  corresponding to the following  form factors:
\begin{align}
	\label{eq:DeVFFL}
	\Delta^{VFF'}_L=&  \frac{1}{16\pi^2 m_V^2} \left\{ g^{LL} m_a \left[  g^{L}_{hFF'}  m_F \left(B^{(1)}_1  +(2m_V^2 +m_{F'}^2 -m_b^2)C_1 \right) 
	\frac{}{}\right. \right. \crn&\left. \left. \hspace{3.1 cm} +g^{R}_{hFF'}  m_{F'} \left( m_V^2 C_0  -B^{(12)}_0 +(2m_V^2 +m_F^2 -m_a^2)C_1 \right)\right]
	\right. \crn&\left. \hspace{1.7cm} +g^{RR} m_b \left[  g^{L}_{hFF'}  m_{F'} \left(  B^{(2)}_1  +(2m_V^2 +m_{F}^2 -m_a^2)C_2 \right) 
	\right. \right. \crn&\left. \left. \hspace{3.3 cm} + g^{R}_{hFF'}  m_{F} \left( m_V^2 C_0- B^{(12)}_0 +(2m_V^2 +m_{F'}^2 -m_b^2)C_2 \right)\right]
	\right. \crn&\left. \hspace{1.7cm} -g^{RL} \left[ g^L_{hFF'}\left( d \times m_V^2 B^{(12)}_0 +4 m_V^2 (m_V^2 C_0 +m_a^2C_1 +m_b^2C_2) 
	\right. \right. \right. \crn&\left. \left.\left.\frac{}{} \hspace{4 cm} -2m_V^2 (m_h^2-m_a^2 -m_b^2) X_0  -A_0(m_V) 
	\frac{}{}\right.\right.\right. \crn &\left. \left.  \left. \frac{}{} \hspace{4cm} - (m_F^2-m_a^2)(B^{(1)}_0 -m_b^2 C_2)  - (m_{F'}^2-m_b^2)(B^{(2)}_0 -m_a^2 C_1)  
	\right. \right. \right. \crn&\left. \left.\left.\frac{}{} \hspace{4 cm} +m_a^2 B^{(1)}_1 +m_b^2 B^{(2)}_1 -(m_F^2-m_a^2)(m_{F'}^2 -m_b^2)C_0  \right) 
	\right.\right. \crn&\left. \left.\hspace{3.cm} + g^R_{hFF'} m_Fm_{F'} \left(3 m_V^2C_0 -B^{(12)}_0\right)
	\right] 
	\right. \crn&\left. \hspace{1.7cm}- g^{LR} g^R_{hFF'} m_a m_b \left[  (m_F^2-m_a^2)  C_1 +(m_{F'}^2 -m_b^2) C_2 -m_V^2 C_0  -B^{(12)}_0 \right] 
	\right\},
	\crn  \Delta^{VFF'}_R=&  \Delta^{VFF'}_L\left[ g^L_{hFF'} \leftrightarrow g^R_{hFF'}, g^{LL}  \leftrightarrow g^{RR},\; g^{LR}  \leftrightarrow g^{RL}\right],
\end{align}
where  $g^{XY}=g^{XY}_{VFF'}$  and the argument of PV-functions is $(m_V^2,m_F^2,m_{F'}^2)$. 

Sum of two diagrams (7) and (8) gives the following form factors:
\begin{align}
	\Delta^{FV}_L=& \frac{g\delta_{hee}}{32\pi^2 m_W m_V^2 (m_a^2 -m_b^2)} 
	\crn &\times \left\{ 
	\left( g^{LL}  m_b +  g^{RR} m_a \right)  m_a m_b\left[2 m_F^2 \left(B^{(2)}_0 -B^{(1)}_0\right) + \left(2 m_V^2  +m_F^2\right) \left( B^{(2)}_1 -B^{(1)}_1\right) 
	\right. \right. \crn &\left. \left. \hspace{5.2cm} +m_b^2 B^{(2)}_1 - m_a^2 B^{(1)}_1\right]  
	\right. \crn &\left. \quad \; + g^{RL}  m_{F} \left[  3m_V^2 \left( m_a^2 B^{(2)}_0 -m_b^2 B^{(1)}_0\right) - \left(m_a^2 -m_b^2\right) A_0(m_F) \right]
	\right. \crn &\left. \quad \; + g^{LR} m_a m_b m_{F} \times  3m_V^2  \left( B^{(2)}_0 -B^{(1)}_0\right)	\right\},
	\crn \Delta^{FV}_R= & \Delta^{FV}_L \left[ g^{LL} \to g^{RR},  g^{RR} \to g^{LL},  g^{RL} \to g^{LR},  g^{LR} \to g^{RL}\right],
\end{align}
where $g^{XY}=g^{XY}_{FVV}$  and   $B^{(k)}_{0,1}=B_{0,1}(p_k^2;m_F^2,m_V^2)$ with $k=1,2$. 

The  form factors corresponding to  diagram $(3)$ in Fig. \ref{fig_ZebaG}  are:
\begin{align}
	\label{eq:DelFVSL}
	\Delta_L^{FVS}= &  \frac{g_{VSh}}{16\pi^2 m_V^2}\left\{m_F \left[ g^{LL} m_a   \left( B^{(1)}_0 +B^{(1)}_1   +(m_V^2 +m_S^2 -m_h^2) C_0  - (m_V^2 -m_S^2 +m_h^2)C_1\right)
	\right. \right. \crn&\left.\left. \hspace{3cm}-g^{RR} m_b \left( 2m_V^2 C_0  + (m_V^2- m_S^2 +m_h^2) C_2 \frac{}{} \right)  \right]  
	\right. \crn & \left. \frac{}{} \hspace{1.5 cm} + g^{RL} \left[  (m_V^2-m_S^2 +m_h^2) \left( B^{(12)}_0 +m_F^2C_0 \right)  -  A_0(m_V) -m_F^2 B^{(1)}_0 - m_a^2 B^{(1)}_1
	\right. \right. \crn&\left.\left. \hspace{2.6 cm} \frac{}{}
	+ \left( m_a^2 (m_V^2 -m_S^2 +m_h^2)  -2 m_V^2  ( m_h^2 -m_b^2)  \right) C_1   + 2m_V^2 m_b^2 C_2  \right] 
	\right. \crn& \left. \frac{}{}  \hspace{1.5 cm}- g^{LR} m_am_b \left[  2m_V^2 C_1 +(m_V^2 +m_S^2 -m_h^2)  C_2 \right]   \right\},
	\crn 	\Delta_R^{FVS}= &	\Delta_L^{FVS}\left[ g^{LL} \to g^{RR},  g^{RR} \to g^{LL},  g^{RL} \to g^{LR},  g^{LR} \to g^{RL}\right],
\end{align}
where $g^{XY}=g^{XY}_{FVS}$and the arguments of PV-functions is $(m_F^2,m_V^2,m_{S}^2)$.

Final results for form factors corresponding to Diagram (4) in Fig. \ref{fig_hebaG}: 
\begin{align}
	\Delta^{FSV}_L= & \frac{g^*_{VSh}}{ 16\pi^2 m_V^2}   \left\{  m_F  \left[ -g^{LL} m_a  \left( 2m_V^2C_0  +(m_V^2 -m_S^2 +m_h^2)C_1 \frac{}{} \right) 
	\right. \right. \crn&\left.  \left. \hspace{2.4cm} +g^{RR}m_b \left( B^{(2)}_0 +B^{(2)}_1 +  (m_V^2 +m_S^2 -m_h^2)C_0 - (m_V^2 -m_S^2 +m_h^2)C_2 \frac{}{}\right) \right]  
	\right. \crn& \left. \hspace{1.6 cm}  +   g^{RL} \left[ (m_V^2 -m_S^2 +m_h^2) ( B^{(12)}_0 +m_F^2 C_0)  - A_0 (m_V) -m_F^2 B^{(2)}_0 -m_b^2 B^{(2)}_1 
	\frac{}{}\right. \right. \crn&\left.  \left. \frac{}{}\hspace{2.6 cm}  +2m_a^2 m_V^2C_1 + \left( m_b^2(m_V^2 - m_S^2 +m_h^2) -2m_V^2(m_h^2 -m_a^2) \right)  C_2  \right]
	\frac{}{}\right. \crn&  \left.\hspace{1.6 cm}  -g^{LR} m_am_b  \left( 2m_V^2C_2 +(m_V^2+m_S^2 -m_h^2) C_1  \frac{}{}\right)   \right\} ,
	\crn 	\Delta^{FSV}_R= &	\Delta^{FSV}_L\left[ g^{LL} \to g^{RR},  g^{RR} \to g^{LL},  g^{RL} \to g^{LR},  g^{LR} \to g^{RL}\right],
\end{align}
where $g^{XY}=g^{XY}_{FSV}$ and the arguments of PV-functions  is $(m_F^2,m_{S}^2,m_V^2)$.

Final results for form factors corresponding to Diagram (5) in Fig. \ref{fig_hebaG}: 
\begin{align}
	\Delta^{FSS'}_L= & \frac{\lambda_{hSS'}}{ 16\pi^2}  \left[ g^{RL}m_F  C_0  -(g^{LL} m_aC_1 +g^{RR} m_bC_2)\right] ,
	\crn 	\Delta^{FSS'}_R= &	\Delta^{FSS'}_L\left[ g^{LL} \to g^{RR},  g^{RR} \to g^{LL},  g^{RL} \to g^{LR},  g^{LR} \to g^{RL}\right],
\end{align}
where  $g^{XY}=g^{XY}_{FSS'}$ and the arguments  $(m_F^2,m_{S}^2,m_{S'}^2)$.

Final results for form factors corresponding to Diagram (6) in Fig. \ref{fig_hebaG}: 
\begin{align}
	\Delta^{SFF'}_L= & \frac{1}{ 16\pi^2}   \left\{  g^L_{hFF'} \left[ g^{RL}   m_F m_{F'} C_0  + g^{RR} m_F  m_b(C_0 +C_2) 
	\frac{}{}\right.\right. \crn& \left. \left. \frac{}{}\hspace{2.2cm} +  g^{LL} m_{F'}  m_a(C_0 +C_1) +g^{LR}  m_a m_b X_0\right]
	\frac{}{} \right. \crn& \left. \frac{}{} \hspace{0.8 cm} + g^R_{hFF'} \left[  g^{RL} \left( B^{(12)}_0 +m_S^2 C_0 +m_a^2 C_1 +m_b^2 C_2 \right) 
	\frac{}{}\right.\right. \crn& \left. \left. \frac{}{}\hspace{2.2cm} + g^{LL} m_a m_F C_1 +g^{RR} m_b m_{F'} C_2 \right]\right\},
	\crn 	\Delta^{SFF'}_R= &	\Delta^{SFF'}_L\left[ g^{LL} \to g^{RR},  g^{RR} \to g^{LL},  g^{RL} \to g^{LR},  g^{LR} \to g^{RL}\right],
\end{align}
where $g^{XY}=g^{XY}_{SFF'}$ and the arguments  $(m_{S}^2,m_F^2,m_{F'}^2)$.

Form factors corresponding to sum of two diagram (9) and (10):
\begin{align}
	\Delta^{FS}_L= &      \frac{g \delta_{hee} }{32\pi^2 m_W(m_a^2 -m_b^2)}   \left[ g^{LR}m_a m_bm_F \left( B^{(1)}_0-B^{(2)}_0 \right)  +g^{RL} m_F \left( m_b^2 B^{(1)}_0 - m_a^2 B^{(2)}_0 \right)  
	\right. \crn & \left.  \hspace{3.5cm} -m_a m_b(g^{LL}  m_b + g^{RR} m_a)  \left( B^{(1)}_1 -B^{(2)}_1 \right)   \frac{}{}\right]   ,
	\crn 	\Delta^{FS}_R= &	\Delta^{FS}_L\left[ g^{LL} \to g^{RR},  g^{RR} \to g^{LL},  g^{RL} \to g^{LR},  g^{LR} \to g^{RL}\right],
\end{align}
where  $g^{XY}=g^{XY}_{FSS}$  and   $B^{(k)}_{0,1}=B_{0,1}(p_k^2; m_F^2,m_S^2)$ with $k=1,2$.

\section{ \label{app:LRIS} Details calculation to the LRIS model}
\subsection{ Higgs sector}

 The model consists of one singly charged Higgs boson, apart from two Goldstone bosons $G^{\pm}_{1,2}$ absorbed by the two respective gauge bosons $W^{\pm}$ and $W'^{\pm}$:
\begin{align}
	\label{eq:CHpm}
	\begin{pmatrix}
		\phi^{\pm}_1	\\
		\phi^{\pm}_2	\\
		\chi_R^{\pm}	
	\end{pmatrix}= & C_{h^\pm}^T\begin{pmatrix}
		G^{\pm}_1	\\
		G^{\pm}_2	\\
		H^{\pm}	
	\end{pmatrix} ,  \; C_{h^\pm}^T = \left(
	\begin{array}{ccc}
		s_{\beta} s_{\xi} & c_{\beta} & c_{\xi} s_{\beta} \\
		c_{\beta} s_{\xi} & -s_{\beta} & c_{\beta} c_{\xi} \\
		c_{\xi} & 0 & -s_{\xi} \\
	\end{array}
	\right), 
\end{align}
where $t_{\xi}\equiv \frac{vc_{2\beta}}{v_R}$, and $C_{h^\pm}$ is the $3\times 3$ unitary matrix satisfying the diagonal relations $C_{h^\pm} \mathcal{M}^2_+C^{T}_{h^\pm} =\mathrm{diag}(0,0,m^2_{H^\pm})$, and  $m^2_{H^\pm}=  \alpha_{32} \left(c_{2\beta} v^2+v_R^2/(2 c_{2\beta})\right)$ with $\alpha_{32}= \alpha_{3} -\alpha_{2}$. 

The CP-odd Higgs components generate two Goldstone bosons of $Z$ and $Z'$, and a physical states $A$. A relation a mong these flavor and mass eigenstates is:
\begin{align}
	\label{eq:CA}
	\begin{pmatrix}
		a_1	\\
		a_2	\\
		a_R	
	\end{pmatrix}= & C_{a}\begin{pmatrix}
		G^{0}_1	\\
		G^{0}_2	\\
		A^0	
	\end{pmatrix} ,  \; C_{a}=\left(
	\begin{array}{ccc}
		0 & -s_{\beta} & c_{\beta} \\
		0 & c_{\beta} & s_{\beta} \\
		1 & 0 & 0 \\
	\end{array}
	\right),
\end{align}
where $m_A^2=2 v^2 (\lambda_3-2 \lambda_2)+ \frac{v_R^2 (\alpha_3-\alpha_2)}{2 c_{2\beta}}$

We will summarize the physical spectrum of the Higgs, boson, and leptons based on previous works \cite{Ezzat:2021bzs}. Some new conventions will be introduced for convenience.

 Regarding to the CP-even Higgs sector. The mass matrix corresponding to the basis $r_H=(r_1,r_2,r_R)^T$, in which $\mathcal{L}_{mass}^H= -\frac{1}{2} r_H^T \mathcal{M}^2_Hr_H$ is:
 \begin{align}
 	\label{eq:M2H}
 	\left(\mathcal{M}^2_H \right)_{11}= & 2 v^2 \left(c_{\beta}^2 \lambda_{23}+2 c_{\beta} \lambda_4 s_{\beta}+\lambda_1 s_{\beta}^2\right) + \frac{\alpha_{32} c_{\beta}^2 v_R^2}{2 \left(c_{\beta}^2-s_{\beta}^2\right)},  
\crn  	\left(\mathcal{M}^2_H \right)_{22}= &2 v^2 \left(c_{\beta}^2 \lambda_1+2 c_{\beta} \lambda_4 s_{\beta}+s_{\beta}^2 \lambda_{23} \right) +\frac{\alpha_{32} s_{\beta}^2 v_R^2}{2 \left(c_{\beta}^2-s_{\beta}^2\right)},
\crn 
\left(\mathcal{M}^2_H \right)_{33}= & 2 \lambda_5 v_R^2,
\crn 
\left(\mathcal{M}^2_H \right)_{12}= & \left(\mathcal{M}^2_H \right)_{21} =2 v^2 \left(c_{\beta}^2 \lambda_4+c_{\beta} s_{\beta} (\lambda_1+\lambda_{23})+\lambda_4 s_{\beta}^2\right)-\frac{\alpha_{32} c_{\beta} s_{\beta} v_R^2}{2 \left(c_{\beta}^2-s_{\beta}^2\right)},
\crn \left(\mathcal{M}^2_H \right)_{13}= &\left(\mathcal{M}^2_H \right)_{31}=  v_R v \left[ \alpha_4 c_{\beta} +s_{\beta}  (\alpha_{12}+\alpha_{32}) \right],
\crn \left(\mathcal{M}^2_H \right)_{23}= & \left(\mathcal{M}^2_H \right)_{32} =v_R v (\alpha_{12} c_{\beta}+\alpha_4 s_{\beta}),
 \end{align}
 where $\lambda_{23}=2\lambda_2 +\lambda_3$.  We see that $\mathrm{Det} \left.\left[ \mathcal{M}^2_H  \right]\right|_{v\to0}=0$, therefore the model consists of at least one neutral CP-even Higgs with mass $\varpropto v^2$, which can be identified with the SM-like Higgs found experimentally. In particular, in the limit $v=0$, the transformation $C_{1}$ can be used to diagonal the squared mass matrix: $C_1\left.\mathcal{M}^2_H  \right|_{v=0} C_1^T= \mathrm{diag} \left(0,\frac{\alpha_{32} v_R^2}{2 c_{2\beta}},2 \lambda_5 v_R^2\right) $, in which 
 \begin{align}
 \label{eq:C1}
 C_1\mathcal{M}^2_H C_1^T &= \mathcal{M}^2_{H,0}, \; C_1= \left(
 \begin{array}{ccc}
 	s_{\beta} & c_{\beta} & 0 \\
 	c_{\beta} & -s_{\beta} & 0 \\
 	0 & 0 & 1 \\
 \end{array}
 \right), 
 \crn  \left(\mathcal{M}^2_{H,0}\right)_{11}&=2 v^2 \left(\lambda_1+\lambda_{23} s_{2\beta}^2 +4 \lambda_4 s_{2\beta}  \right),
 \crn \left(\mathcal{M}^2_{H,0}\right)_{22}&=2 c_{2\beta}^2 \lambda_{23} v^2+\frac{\alpha_{32} v_R^2}{2 c_{2\beta}},
 \crn \left(\mathcal{M}^2_{H,0}\right)_{33}&= 2 \lambda_5 v_R^2,
 \crn \left(\mathcal{M}^2_{H,0}\right)_{12}&=\left(\mathcal{M}^2_{H,0}\right)_{21} = 2 c_{2\beta} v^2 (\lambda_4+\lambda_{23} s_{2\beta}),
 \crn \left(\mathcal{M}^2_{H,0}\right)_{13}&= \left(\mathcal{M}^2_{H,0}\right)_{31}= v_R v \left(\alpha_{12}+\alpha_4 s_{2\beta}+\alpha_{32} s_{\beta}^2\right),
 \crn \left(\mathcal{M}^2_{H,0}\right)_{23}&= \left(\mathcal{M}^2_{H,0}\right)_{32}= v_R v \left(\alpha_4 c_{2\beta}+\frac{\alpha_{32} s_{2\beta}}{2}\right).
 \end{align}
 It can be seen that $\left(\mathcal{M}^2_{H,0}\right)_{11}\varpropto v^2 \ll v_R^2 \varpropto \left(\mathcal{M}^2_{H,0}\right)_{22}, \left(\mathcal{M}^2_{H,0}\right)_{33}$. In addition, non-diagonal entries of  $\mathcal{M}^2_{H,0} \varpropto vv_R\ll \left(\mathcal{M}^2_{H,0}\right)_{22}, \left(\mathcal{M}^2_{H,0}\right)_{33}$, therefore, the mixing matrix used to diagonalized $\mathcal{M}^2_{H,0}$ is close to identity. A a result, we will use $C_1$ as the relation between the flavor states and the physical states $(h,h_1,h_2)$, in which $h$ is identified with the SM-like Higgs boson: $(r_1,r_2^T,r_R)^T=C_1^T(h,h_1,h_2)^T$. 
 This corresponds to the strict relations of $\left(\mathcal{M}^2_{H,0}\right)_{12}=\left(\mathcal{M}^2_{H,0}\right)_{13}=\left(\mathcal{M}^2_{H,0}\right)_{23}=0$, equivalently $\lambda_4=-\lambda_{23} s_{2\beta}$, $\alpha_{12}= -\alpha_4 s_{2\beta} -\alpha_{32} s_{\beta}^2$, and $\alpha_4  =-\frac{\alpha_{32} t_{2\beta}}{2}$. As a result, $m_h^2=2 v^2 \left(\lambda_1+\lambda_{23} s_{2\beta}^2 +4 \lambda_4 s_{2\beta}  \right)$. 
 
\subsection{ \label{subsec:LFV} One-loop contributions to decays LFV$h$ and LFV$Z$}

We list here the one-loop contributions from charge Higgs and gauge bosons $H^\pm$, $W^\pm$ and $W'^\pm$, which contain large contributions to $\Delta a_{e_a}$ and LFV decay rates. 

For LFV$h$ decays, one-loop contributions are collected as follows.  Diagram (1) of Fig. \ref{fig_hebaG} give:
\begin{align}
\label{eq:DeltaLRFVVh}
%
\Delta^{(1+2+7+8)}_{L,R} =& \sum_{i=1}^9 \left[ \Delta^{ iWW}_{L,R}  + \Delta^{iW'W'}_{L,R}+ \Delta^{iWW'}_{L,R} +\Delta^{iW'W}_{L,R}    + \Delta^{FW}_{L,R} + \Delta^{FW'}_{L,R}\right]
\crn&+ \sum_{i,j=1}^9\left[  \Delta^{Wij}_{L,R} +\Delta^{W'ij}_{L,R}\right],
\end{align} 
where $g_{hVV'}=g_{hWW}, g_{hW'W'}, g_{hWW'}$, and $g_{hWW'}$ were given in Table \ref{t:hXX};  and $\delta_{hee}=m_W/(gv/2)$ derived from Eq. \eqref{eq:LllH}. 
Diagram from singly charged Higgs boson $H^\pm$:
\begin{align}
\label{eq_DeltaLRFSSh}
 \Delta^{(5+6+9+10)}_{L,R} &=  \sum_{i=1}^9 \left[ \Delta^{iH^+H^+}_{L,R}  + \Delta^{ iH^+}_{L,R}\right] +\sum_{i,j=1}^9 \Delta^{ H^+ij}_{L,R},
\end{align}

Diagram from both exchanges of singly charged Higgs boson $H^\pm$ and gauge boson:
\begin{align}
	\label{eq_DeltaLRFVS}
\Delta^{(3+4)}_{L,R} &= \sum_{i=1}^9 \left( \Delta^{iH^+W}_{L,R} +\Delta^{ iWH^+}_{L,R} + \Delta^{iH^+W'}_{L,R} +\Delta^{ iW'H^+}_{L,R}\right),
\end{align}
where $g_{VSh}=g_{W^+H^-h}, g_{W'^+H^-h}$ are given in Table \ref{t:hXX}.

The one-loop contributions to LFV$Z$ decay rates presented in Fig. \ref{fig_ZebaG} are collected as follows.  The total form factors are sum of all particular contributions as follows
\begin{align}
	\label{eq:ZFVV}
	\bar{a}_{L,R}&= \bar{a}^{(1+2+7+8)}_{L,R} +\bar{a}^{(5+6+9+10)}_{L,R} +\bar{a}^{(3+4)}_{L,R},  
	\crn  \bar{b}_{L,R} &= \bar{b}^{(1+2+7+8)}_{L,R} +\bar{b}^{(5+6+9+10)}_{L,R} +\bar{b}^{(3+4)}_{L,R}, 
	\crn 	\bar{a}^{(1+2+7+8)}_{L,R} =& \sum_{i=1}^9 \left[ \bar{a}^{ iWW}_{L,R}  + \bar{a}^{iW'W'}_{L,R}+ \bar{a}^{iWW'}_{L,R} +\bar{a}^{iW'W}_{L,R}    +\bar{a}^{iW}_{L,R} + \bar{a}^{iW'}_{L,R}\right]
\crn&+ \sum_{i,j=1}^9\left[  \bar{a}^{Wij}_{L,R} +\bar{a}^{W'ij}_{L,R}\right],
\crn 	\bar{b}^{(1+2+7+8)}_{L,R} =& \sum_{i=1}^9 \left[ \bar{b}^{ iWW}_{L,R}  +\bar{b}^{iW'W'}_{L,R}+ \bar{b}^{iWW'}_{L,R} +\bar{b}^{iW'W}_{L,R}    + \bar{b}^{iW}_{L,R} + \bar{b}^{FW'}_{L,R}\right]
\crn&+ \sum_{i,j=1}^9\left[  \bar{b}^{Wij}_{L,R} +\bar{b}^{W'ij}_{L,R}\right],  
\end{align}

From singly charged Higgs boson:
\begin{align}
	\label{eq:ZFSS}
\bar{a}^{(5+6+9+10)}_{L,R}  =&\sum_{i=1}^9 \left[ \bar{a}^{iH^+H^-}_{L,R}  + \bar{a}^{ iH^+}_{L,R}\right] +\sum_{i,j=1}^9\bar{a}^{ H^+ij}_{L,R},
	\crn 	\bar{b}^{(5+6+9+10)}_{L,R}  =&\sum_{i=1}^9 \left[ \bar{b}^{iH^+H^+}_{L,R}  + \bar{b}^{ iH^+}_{L,R}\right] +\sum_{i,j=1}^9 \bar{b}^{ H^+ij}_{L,R}, 
\end{align}

From both Higgs and gauge boson exchanges in diagram (3+4)
\begin{align}
	\label{eq:ZFSV}
	\bar{a}^{(3+4)}_{L,R} &= \sum_{i=1}^9 \left( \bar{a}^{iH^+W}_{L,R} +\bar{a}^{ iWH^+}_{L,R} + \bar{a}^{iH^+W'}_{L,R} +\bar{a}^{ iW'H^+}_{L,R}\right),
	\crn \bar{b}^{ (3+4)}_{L,R} &=\sum_{i=1}^9 \left( \bar{b}^{iH^+W}_{L,R} +\bar{b}^{ iWH^+}_{L,R} + \bar{b}^{iH^+W'}_{L,R} +\bar{b}^{ iW'H^+}_{L,R}\right),
	\end{align}
where $g_{SVZ}=g_{H^-W^+Z}, g_{H^-W'^+Z}$ are given in Table \ref{table:Zxx}.


\begin{thebibliography}{99}
\bibitem{CMS:2021rsq}
A.~M.~Sirunyan \textit{et al.} [CMS],
Phys. Rev. D \textbf{104} (2021) no.3, 032013 
[arXiv:2105.03007 [hep-ex]].

\bibitem{ATLAS:2019xlq}
[ATLAS],
``Search for the decays of the Higgs boson $H \to ee$ and $H \to e\mu$ in $pp$ collisions at $\sqrt{s}$ = 13 TeV with the ATLAS detector,''
ATLAS-CONF-2019-037

\bibitem{CMS:2023pte}
A.~Hayrapetyan \textit{et al.} [CMS],
Phys. Rev. D \textbf{108} (2023) no.7, 072004
[arXiv:2305.18106 [hep-ex]].

\bibitem{ATLAS:2023mvd}
G.~Aad \textit{et al.} [ATLAS],
JHEP \textbf{07} (2023), 166
[arXiv:2302.05225 [hep-ex]].

\bibitem{ATLAS:2021bdj}
G.~Aad \textit{et al.} [ATLAS],
Phys. Rev. Lett. \textbf{127} (2022), 271801
[arXiv:2105.12491 [hep-ex]].

\bibitem{ATLAS:2022uhq}
G.~Aad \textit{et al.} [ATLAS],
Phys. Rev. D \textbf{108} (2023), 032015
[arXiv:2204.10783 [hep-ex]].

\bibitem{Pilaftsis:1992st}
A.~Pilaftsis,
Phys. Lett. B \textbf{285} (1992), 68-74

\bibitem{Korner:1992zk}
J.~G.~Korner, A.~Pilaftsis and K.~Schilcher,
Phys. Rev. D \textbf{47} (1993), 1080-1086
[arXiv:hep-ph/9301289 [hep-ph]].

\bibitem{Lavoura:2003xp}
L.~Lavoura,
Eur. Phys. J. C \textbf{29} (2003), 191-195
[arXiv:hep-ph/0302221 [hep-ph]].

\bibitem{Crivellin:2018qmi}
A.~Crivellin, M.~Hoferichter and P.~Schmidt-Wellenburg,
Phys. Rev. D \textbf{98} (2018) no.11, 113002
[arXiv:1807.11484 [hep-ph]].



\bibitem{Diaz-Cruz:1999sns}
J.~L.~Diaz-Cruz and J.~J.~Toscano,
Phys. Rev. D \textbf{62} (2000), 116005
[arXiv:hep-ph/9910233 [hep-ph]].

\bibitem{Arganda:2004bz}
E.~Arganda, A.~M.~Curiel, M.~J.~Herrero and D.~Temes,
Phys. Rev. D \textbf{71} (2005), 035011
[arXiv:hep-ph/0407302 [hep-ph]].

\bibitem{Arganda:2014dta}
E.~Arganda, M.~J.~Herrero, X.~Marcano and C.~Weiland,
Phys. Rev. D \textbf{91} (2015) no.1, 015001
[arXiv:1405.4300 [hep-ph]].

\bibitem{Hue:2015fbb}
L.~T.~Hue, H.~N.~Long, T.~T.~Thuc and T.~Phong Nguyen,
Nucl. Phys. B \textbf{907} (2016), 37-76
[arXiv:1512.03266 [hep-ph]].

\bibitem{Chiang:2016vgf}
C.~W.~Chiang, K.~Fuyuto and E.~Senaha,
Phys. Lett. B \textbf{762} (2016), 315-320
[arXiv:1607.07316 [hep-ph]].


\bibitem{Arganda:2015uca}
E.~Arganda, M.~J.~Herrero, R.~Morales and A.~Szynkman,
JHEP \textbf{03} (2016), 055
[arXiv:1510.04685 [hep-ph]].

\bibitem{Baek:2015mea}
S.~Baek and K.~Nishiwaki,
Phys. Rev. D \textbf{93} (2016) no.1, 015002
[arXiv:1509.07410 [hep-ph]].

\bibitem{Arganda:2015naa}
E.~Arganda, M.~J.~Herrero, X.~Marcano and C.~Weiland,
Phys. Rev. D \textbf{93} (2016) no.5, 055010
[arXiv:1508.04623 [hep-ph]].


\bibitem{Thuc:2016qva}
T.~T.~Thuc, L.~T.~Hue, H.~N.~Long and T.~P.~Nguyen,
Phys. Rev. D \textbf{93} (2016) no.11, 115026
[arXiv:1604.03285 [hep-ph]].

\bibitem{Arganda:2016zvc}
E.~Arganda, M.~J.~Herrero, X.~Marcano, R.~Morales and A.~Szynkman,
Phys. Rev. D \textbf{95} (2017) no.9, 095029
[arXiv:1612.09290 [hep-ph]].


\bibitem{Cai:2017jrq}
Y.~Cai, J.~Herrero-Garc\'\i{}a, M.~A.~Schmidt, A.~Vicente and R.~R.~Volkas,
Front. in Phys. \textbf{5} (2017), 63
[arXiv:1706.08524 [hep-ph]].



\bibitem{Nguyen:2018rlb}
T.~P.~Nguyen, T.~T.~Le, T.~T.~Hong and L.~T.~Hue,
Phys. Rev. D \textbf{97} (2018) no.7, 073003
[arXiv:1802.00429 [hep-ph]].

\bibitem{Zhang:2021nzv}
Z.~N.~Zhang, H.~B.~Zhang, J.~L.~Yang, S.~M.~Zhao and T.~F.~Feng,
Phys. Rev. D \textbf{103} (2021) no.11, 115015
[arXiv:2105.09799 [hep-ph]].

\bibitem{Hong:2022xjg}
T.~T.~Hong, N.~H.~T.~Nha, T.~P.~Nguyen, L.~T.~T.~Phuong and L.~T.~Hue,
PTEP \textbf{2022} (2022) no.9, 093B05
[arXiv:2206.08028 [hep-ph]].

\bibitem{Nguyen:2020ehj}
T.~P.~Nguyen, T.~T.~Thuc, D.~T.~Si, T.~T.~Hong and L.~T.~Hue,
PTEP \textbf{2022} (2022) no.2, 023B01

\bibitem{CarcamoHernandez:2020pnh}
A.~E.~C\'arcamo Hern\'andez, L.~T.~Hue, S.~Kovalenko and H.~N.~Long,
Eur. Phys. J. Plus \textbf{136} (2021) no.11, 1158
[arXiv:2001.01748 [hep-ph]].

\bibitem{Marcano:2019rmk}
X.~Marcano and R.~A.~Morales,
Front. in Phys. \textbf{7} (2020), 228
[arXiv:1909.05888 [hep-ph]].


\bibitem{Chen:2023eof}
C.~H.~Chen, C.~W.~Chiang and C.~W.~Su,
J. Phys. G \textbf{51} (2024) no.8, 085001
[arXiv:2301.07070 [hep-ph]].


\bibitem{Korner:1992an}
J.~G.~Korner, A.~Pilaftsis and K.~Schilcher,
Phys. Lett. B \textbf{300} (1993), 381-386
[arXiv:hep-ph/9301290 [hep-ph]].
%
\bibitem{DeRomeri:2016gum}
V.~De Romeri, M.~J.~Herrero, X.~Marcano and F.~Scarcella,
Phys. Rev. D \textbf{95} (2017) no.7, 075028
[arXiv:1607.05257 [hep-ph]].
\bibitem{Abada:2021zcm}
A.~Abada, J.~Kriewald and A.~M.~Teixeira,
Eur. Phys. J. C \textbf{81} (2021) no.11, 1016
[arXiv:2107.06313 [hep-ph]].

\bibitem{Hernandez-Tome:2019lkb}
G.~Hern\'andez-Tom\'e, J.~I.~Illana, M.~Masip, G.~L\'opez Castro and P.~Roig,
Phys. Rev. D \textbf{101} (2020) no.7, 075020
[arXiv:1912.13327 [hep-ph]].

\bibitem{Crivellin:2018mqz}
A.~Crivellin, Z.~Fabisiewicz, W.~Materkowska, U.~Nierste, S.~Pokorski and J.~Rosiek,
JHEP \textbf{06} (2018), 003
[arXiv:1802.06803 [hep-ph]].

\bibitem{Ilakovac:2012sh}
A.~Ilakovac, A.~Pilaftsis and L.~Popov,
Phys. Rev. D \textbf{87} (2013) no.5, 053014
[arXiv:1212.5939 [hep-ph]].

\bibitem{Herrero:2018luu}
M.~J.~Herrero, X.~Marcano, R.~Morales and A.~Szynkman,
Eur. Phys. J. C \textbf{78} (2018) no.10, 815
[arXiv:1807.01698 [hep-ph]].

\bibitem{Calibbi:2021pyh}
L.~Calibbi, X.~Marcano and J.~Roy,
Eur. Phys. J. C \textbf{81} (2021) no.12, 1054
[arXiv:2107.10273 [hep-ph]]

\bibitem{Jurciukonis:2021izn}
D.~Jur\v{c}iukonis and L.~Lavoura,
JHEP \textbf{03} (2022), 106
[arXiv:2107.14207 [hep-ph]].

\bibitem{Abada:2022asx}
A.~Abada, J.~Kriewald, E.~Pinsard, S.~Rosauro-Alcaraz and A.~M.~Teixeira,
Eur. Phys. J. C \textbf{83} (2023) no.6, 494
[arXiv:2207.10109 [hep-ph]].

\bibitem{Crivellin:2022cve}
A.~Crivellin, F.~Kirk and C.~A.~Manzari,
JHEP \textbf{12} (2022), 031
[arXiv:2208.00020 [hep-ph]].

\bibitem{Hong:2023rhg}
T.~T.~Hong, Q.~D.~Tran, T.~P.~Nguyen, L.~T.~Hue and N.~H.~T.~Nha,
Eur. Phys. J. C \textbf{84} (2024) no.3, 338
[erratum: Eur. Phys. J. C \textbf{84} (2024) no.5, 454]
[arXiv:2312.11427 [hep-ph]].




\bibitem{Hong:2024yhk}
T.~T.~Hong, L.~T.~T.~Phuong, T.~P.~Nguyen, N.~H.~T.~Nha and L.~T.~Hue,
Phys. Rev. D \textbf{110} (2024) no.7, 075010
[arXiv:2404.05524 [hep-ph]].

\bibitem{Hong:2024swk}
T.~T.~Hong, L.~T.~Hue, L.~T.~T.~Phuong, N.~H.~T.~Nha and T.~P.~Nguyen,
Phys. Scripta \textbf{99} (2024) no.12, 125308
[arXiv:2406.11040 [hep-ph]].

\bibitem{Kriewald:2022erk}
J.~Kriewald, J.~Orloff, E.~Pinsard and A.~M.~Teixeira,
Eur. Phys. J. C \textbf{82} (2022) no.9, 844
[arXiv:2204.13134 [hep-ph]].

\bibitem{Pati:1974yy}
J.~C.~Pati and A.~Salam,
Phys. Rev. D \textbf{10} (1974), 275-289
[erratum: Phys. Rev. D \textbf{11} (1975), 703-703]


\bibitem{Mohapatra:1974gc}
R.~N.~Mohapatra and J.~C.~Pati,
Phys. Rev. D \textbf{11} (1975), 2558


\bibitem{Mohapatra:1974hk}
R.~N.~Mohapatra and J.~C.~Pati,
Phys. Rev. D \textbf{11} (1975), 566-571

\bibitem{Senjanovic:1975rk}
G.~Senjanovic and R.~N.~Mohapatra,
Phys. Rev. D \textbf{12} (1975), 1502


\bibitem{Senjanovic:1978ev}
G.~Senjanovic,
Nucl. Phys. B \textbf{153} (1979), 334-364

\bibitem{Mohapatra:1979ia}
R.~N.~Mohapatra and G.~Senjanovic,
Phys. Rev. Lett. \textbf{44} (1980), 912

\bibitem{ThomasArun:2021rwf}
M.~Thomas Arun, T.~Mandal, S.~Mitra, A.~Mukherjee, L.~Priya and A.~Sampath,
Phys. Rev. D \textbf{105} (2022) no.11, 115007
[arXiv:2109.09585 [hep-ph]].
\bibitem{Hue:2017lak}
L.~T.~Hue, L.~D.~Ninh, T.~T.~Thuc and N.~T.~T.~Dat,
Eur. Phys. J. C \textbf{78} (2018) no.2, 128
[arXiv:1708.09723 [hep-ph]].

\bibitem{Hue:2023rks}
L.~T.~Hue, H.~N.~Long, V.~H.~Binh, H.~L.~T.~Mai and T.~P.~Nguyen,
Nucl. Phys. B \textbf{992} (2023), 116244
[arXiv:2301.05407 [hep-ph]]. 

\bibitem{tHooft:1978jhc}
G.~'t Hooft and M.~J.~G.~Veltman,
Nucl. Phys. B \textbf{153} (1979), 365-401

\bibitem{Hahn:1998yk}
T.~Hahn and M.~Perez-Victoria,
Comput. Phys. Commun. \textbf{118} (1999), 153-165
[arXiv:hep-ph/9807565 [hep-ph]].

\bibitem{Super-Kamiokande:1998kpq}
Y.~Fukuda \textit{et al.} [Super-Kamiokande],
Phys. Rev. Lett. \textbf{81} (1998), 1562-1567
[arXiv:hep-ex/9807003 [hep-ex]].

\bibitem{Super-Kamiokande:2000ywb}
S.~Fukuda \textit{et al.} [Super-Kamiokande],
Phys. Rev. Lett. \textbf{85} (2000), 3999-4003
[arXiv:hep-ex/0009001 [hep-ex]].

\bibitem{SNO:2002tuh}
Q.~R.~Ahmad \textit{et al.} [SNO],
Phys. Rev. Lett. \textbf{89} (2002), 011301
[arXiv:nucl-ex/0204008 [nucl-ex]].

\bibitem{Li:2022zap}
S.~Li, Z.~Li, F.~Wang and J.~M.~Yang,
Nucl. Phys. B \textbf{983} (2022), 115927
[arXiv:2205.15153 [hep-ph]].


\bibitem{CarcamoHernandez:2024edi}
A.~E.~C\'arcamo Hern\'andez, Y.~H.~Vel\'asquez, S.~Kovalenko, N.~A.~P\'erez-Julve and I.~Schmidt,
[arXiv:2403.05637 [hep-ph]].

\bibitem{Escribano:2021css}
P.~Escribano, J.~Terol-Calvo and A.~Vicente,
Phys. Rev. D \textbf{103} (2021) no.11, 115018
[arXiv:2104.03705 [hep-ph]].


\bibitem{Hue:2021xap}
L.~T.~Hue, H.~T.~Hung, N.~T.~Tham, H.~N.~Long and T.~P.~Nguyen,
Phys. Rev. D \textbf{104} (2021) no.3, 033007
[arXiv:2104.01840 [hep-ph]].



\bibitem{Vermaseren:2000nd}
J.~A.~M.~Vermaseren,
[arXiv:math-ph/0010025 [math-ph]].

\bibitem{Phan:2016ouz}
K.~H.~Phan, H.~T.~Hung and L.~T.~Hue,
PTEP \textbf{2016} (2016) no.11, 113B03




\bibitem{Zeleny-Mora:2021tym}
M.~Zeleny-Mora, J.~L.~D\'\i{}az-Cruz and O.~F\'elix-Beltr\'an,
Int. J. Mod. Phys. A \textbf{37} (2022) no.36, 2250226
[arXiv:2112.08412 [hep-ph]].

\bibitem{LHCHiggsCrossSectionWorkingGroup:2016ypw}
D.~de Florian \textit{et al.} [LHC Higgs Cross Section Working Group],
[arXiv:1610.07922 [hep-ph]].

\bibitem{Dreiner:2008tw}
H.~K.~Dreiner, H.~E.~Haber and S.~P.~Martin,
Phys. Rept. \textbf{494} (2010), 1-196
[arXiv:0812.1594 [hep-ph]].


\bibitem{Ghorbani:2022muk}
P.~Ghorbani,
JHEP \textbf{04} (2022), 170
[arXiv:2203.03964 [hep-ph]].


\bibitem{Bhaskar:2022vgk}
A.~Bhaskar, A.~A.~Madathil, T.~Mandal and S.~Mitra,
Phys. Rev. D \textbf{106} (2022) no.11, 115009
[arXiv:2204.09031 [hep-ph]].



\bibitem{Dev:2016dja}
P.~S.~B.~Dev, R.~N.~Mohapatra and Y.~Zhang,
JHEP \textbf{05} (2016), 174
[arXiv:1602.05947 [hep-ph]].

\bibitem{Khalil:2010iu}
S.~Khalil,
Phys. Rev. D \textbf{82} (2010), 077702
[arXiv:1004.0013 [hep-ph]].

\bibitem{Ezzat:2021bzs}
K.~Ezzat, M.~Ashry and S.~Khalil,
Phys. Rev. D \textbf{104} (2021) no.1, 015016
[arXiv:2101.08255 [hep-ph]].

\bibitem{Ezzat:2022gpk}
K.~Ezzat, G.~Faisel and S.~Khalil,
Eur. Phys. J. C \textbf{83} (2023) no.8, 731
[arXiv:2204.10922 [hep-ph]].

\bibitem{Ashry:2022maw}
M.~Ashry, K.~Ezzat and S.~Khalil,
Phys. Rev. D \textbf{107} (2023) no.5, 055044
[arXiv:2207.05828 [hep-ph]].


\bibitem{Maki:1962mu}
Z.~Maki, M.~Nakagawa and S.~Sakata,
Prog. Theor. Phys. \textbf{28} (1962), 870-880

\bibitem{Pontecorvo:1967fh}
B.~Pontecorvo,
Zh. Eksp. Teor. Fiz. \textbf{53} (1967), 1717-1725

\bibitem{Workman:2022ynf}
R.~L.~Workman [Particle Data Group],
PTEP \textbf{2022}, 083C01 (2022)

\bibitem{Zhang:2007da}
Y.~Zhang, H.~An, X.~Ji and R.~N.~Mohapatra,
Nucl. Phys. B \textbf{802} (2008), 247-279
[arXiv:0712.4218 [hep-ph]].


\bibitem{Planck:2018vyg}
N.~Aghanim \textit{et al.} [Planck],
Astron. Astrophys. \textbf{641} (2020), A6
[erratum: Astron. Astrophys. \textbf{652} (2021), C4]
[arXiv:1807.06209 [astro-ph.CO]].


\bibitem{Muong-2:2023cdq}
D.~P.~Aguillard \textit{et al.} [Muon g-2],
Phys. Rev. Lett. \textbf{131} (2023) no.16, 161802
[arXiv:2308.06230 [hep-ex]].

\bibitem{Aoyama:2020ynm}
T.~Aoyama, N.~Asmussen, M.~Benayoun, J.~Bijnens, T.~Blum, M.~Bruno, I.~Caprini, C.~M.~Carloni Calame, M.~C\`e and G.~Colangelo, \textit{et al.}
Phys. Rept. \textbf{887}, 1-166 (2020)
[arXiv:2006.04822 [hep-ph]].

\bibitem{Davier:2010nc}
M.~Davier, A.~Hoecker, B.~Malaescu and Z.~Zhang,
Eur. Phys. J. C \textbf{71}, 1515 (2011)
[erratum: Eur. Phys. J. C \textbf{72}, 1874 (2012)]
[arXiv:1010.4180 [hep-ph]].

\bibitem{Aoyama:2012wk}
T.~Aoyama, M.~Hayakawa, T.~Kinoshita and M.~Nio,
Phys. Rev. Lett. \textbf{109}, 111808 (2012)
[arXiv:1205.5370 [hep-ph]].

\bibitem{Aoyama:2019ryr}
T.~Aoyama, T.~Kinoshita and M.~Nio,
Atoms \textbf{7}, no.1, 28 (2019)

\bibitem{Czarnecki:2002nt}
A.~Czarnecki, W.~J.~Marciano and A.~Vainshtein,
Phys. Rev. D \textbf{67}, 073006 (2003)
[erratum: Phys. Rev. D \textbf{73}, 119901 (2006)]
[arXiv:hep-ph/0212229 [hep-ph]].

\bibitem{Gnendiger:2013pva}
C.~Gnendiger, D.~St\"ockinger and H.~St\"ockinger-Kim,
Phys. Rev. D \textbf{88}, 053005 (2013)
[arXiv:1306.5546 [hep-ph]].

\bibitem{Danilkin:2016hnh}
I.~Danilkin and M.~Vanderhaeghen,
Phys. Rev. D \textbf{95}, no.1, 014019 (2017)
[arXiv:1611.04646 [hep-ph]].

\bibitem{Davier:2017zfy}
M.~Davier, A.~Hoecker, B.~Malaescu and Z.~Zhang,
Eur. Phys. J. C \textbf{77}, no.12, 827 (2017)
[arXiv:1706.09436 [hep-ph]].

\bibitem{Keshavarzi:2018mgv}
A.~Keshavarzi, D.~Nomura and T.~Teubner,
Phys. Rev. D \textbf{97}, no.11, 114025 (2018)
[arXiv:1802.02995 [hep-ph]].

\bibitem{Colangelo:2018mtw}
G.~Colangelo, M.~Hoferichter and P.~Stoffer,
JHEP \textbf{02}, 006 (2019)
[arXiv:1810.00007 [hep-ph]].

\bibitem{Hoferichter:2019mqg}
M.~Hoferichter, B.~L.~Hoid and B.~Kubis,
JHEP \textbf{08}, 137 (2019)
[arXiv:1907.01556 [hep-ph]].

\bibitem{Davier:2019can}
M.~Davier, A.~Hoecker, B.~Malaescu and Z.~Zhang,
Eur. Phys. J. C \textbf{80}, no.3, 241 (2020)
[erratum: Eur. Phys. J. C \textbf{80}, no.5, 410 (2020)]
[arXiv:1908.00921 [hep-ph]].

\bibitem{Keshavarzi:2019abf}
A.~Keshavarzi, D.~Nomura and T.~Teubner,
Phys. Rev. D \textbf{101}, no.1, 014029 (2020)
[arXiv:1911.00367 [hep-ph]].

\bibitem{Kurz:2014wya}
A.~Kurz, T.~Liu, P.~Marquard and M.~Steinhauser,
Phys. Lett. B \textbf{734}, 144-147 (2014)
[arXiv:1403.6400 [hep-ph]].

\bibitem{Melnikov:2003xd}
K.~Melnikov and A.~Vainshtein,
Phys. Rev. D \textbf{70}, 113006 (2004)
[arXiv:hep-ph/0312226 [hep-ph]].

\bibitem{Masjuan:2017tvw}
P.~Masjuan and P.~Sanchez-Puertas,
Phys. Rev. D \textbf{95}, no.5, 054026 (2017)
[arXiv:1701.05829 [hep-ph]].

\bibitem{Colangelo:2017fiz}
G.~Colangelo, M.~Hoferichter, M.~Procura and P.~Stoffer,
JHEP \textbf{04}, 161 (2017)
[arXiv:1702.07347 [hep-ph]].

\bibitem{Hoferichter:2018kwz}
M.~Hoferichter, B.~L.~Hoid, B.~Kubis, S.~Leupold and S.~P.~Schneider,
JHEP \textbf{10}, 141 (2018)
[arXiv:1808.04823 [hep-ph]].

\bibitem{Gerardin:2019vio}
A.~G\'erardin, H.~B.~Meyer and A.~Nyffeler,
Phys. Rev. D \textbf{100}, no.3, 034520 (2019)
[arXiv:1903.09471 [hep-lat]].

\bibitem{Bijnens:2019ghy}
J.~Bijnens, N.~Hermansson-Truedsson and A.~Rodr\'\i{}guez-S\'anchez,
Phys. Lett. B \textbf{798}, 134994 (2019)
[arXiv:1908.03331 [hep-ph]].

\bibitem{Colangelo:2019uex}
G.~Colangelo, F.~Hagelstein, M.~Hoferichter, L.~Laub and P.~Stoffer,
JHEP \textbf{03}, 101 (2020)
[arXiv:1910.13432 [hep-ph]].

\bibitem{Blum:2019ugy}
T.~Blum, N.~Christ, M.~Hayakawa, T.~Izubuchi, L.~Jin, C.~Jung and C.~Lehner,
Phys. Rev. Lett. \textbf{124}, no.13, 132002 (2020)
[arXiv:1911.08123 [hep-lat]].

\bibitem{Colangelo:2014qya}
G.~Colangelo, M.~Hoferichter, A.~Nyffeler, M.~Passera and P.~Stoffer,
Phys. Lett. B \textbf{735}, 90-91 (2014)
[arXiv:1403.7512 [hep-ph]].
\bibitem{Pauk:2014rta}
V.~Pauk and M.~Vanderhaeghen,
Eur. Phys. J. C \textbf{74}, no.8, 3008 (2014)
[arXiv:1401.0832 [hep-ph]].

\bibitem{Jegerlehner:2017gek}
F.~Jegerlehner,
Springer Tracts Mod. Phys. \textbf{274}, pp.1-693 (2017)

\bibitem{Knecht:2018sci}
M.~Knecht, S.~Narison, A.~Rabemananjara and D.~Rabetiarivony,
Phys. Lett. B \textbf{787}, 111-123 (2018)
[arXiv:1808.03848 [hep-ph]].

\bibitem{Eichmann:2019bqf}
G.~Eichmann, C.~S.~Fischer and R.~Williams,
Phys. Rev. D \textbf{101}, no.5, 054015 (2020)
[arXiv:1910.06795 [hep-ph]].

\bibitem{Roig:2019reh}
P.~Roig and P.~Sanchez-Puertas,
Phys. Rev. D \textbf{101}, no.7, 074019 (2020)
[arXiv:1910.02881 [hep-ph]].

\bibitem{Venanzoni:2023mbe}
G.~Venanzoni [Muon g-2],
PoS \textbf{EPS-HEP2023} (2024), 037
[arXiv:2311.08282 [hep-ex]].


\bibitem{Hanneke:2008tm}
D.~Hanneke, S.~Fogwell and G.~Gabrielse,
Phys. Rev. Lett. \textbf{100} (2008), 120801
[arXiv:0801.1134 [physics.atom-ph]].

\bibitem{Parker:2018vye}
R.~H.~Parker, C.~Yu, W.~Zhong, B.~Estey and H.~M\"uller,
Science \textbf{360} (2018), 191
[arXiv:1812.04130 [physics.atom-ph]].

\bibitem{Fan:2022eto}
X.~Fan, T.~G.~Myers, B.~A.~D.~Sukra and G.~Gabrielse,
Phys. Rev. Lett. \textbf{130} (2023) no.7, 071801
[arXiv:2209.13084 [physics.atom-ph]].

\bibitem{Morel:2020dww}
L.~Morel, Z.~Yao, P.~Clad\'e and S.~Guellati-Kh\'elifa,
Nature \textbf{588} (2020) no.7836, 61-65

\bibitem{Aoyama:2012wj}
T.~Aoyama, M.~Hayakawa, T.~Kinoshita and M.~Nio,
Phys. Rev. Lett. \textbf{109} (2012), 111807
[arXiv:1205.5368 [hep-ph]].



\bibitem{Laporta:2017okg}
S.~Laporta,
Phys. Lett. B \textbf{772} (2017), 232-238
[arXiv:1704.06996 [hep-ph]].

\bibitem{Aoyama:2017uqe}
T.~Aoyama, T.~Kinoshita and M.~Nio,
Phys. Rev. D \textbf{97} (2018) no.3, 036001
[arXiv:1712.06060 [hep-ph]].

\bibitem{Terazawa:2018pdc}
H.~Terazawa,
Nonlin. Phenom. Complex Syst. \textbf{21} (2018) no.3, 268-272

\bibitem{Volkov:2019phy}
S.~Volkov,
Phys. Rev. D \textbf{100} (2019) no.9, 096004
[arXiv:1909.08015 [hep-ph]].



\bibitem{Gerardin:2020gpp}
A.~G\'erardin,
Eur. Phys. J. A \textbf{57} (2021) no.4, 116
[arXiv:2012.03931 [hep-lat]].

\bibitem{BaBar:2009hkt}
B.~Aubert \textit{et al.} [BaBar],
Phys. Rev. Lett. \textbf{104} (2010), 021802
[arXiv:0908.2381 [hep-ex]].


\bibitem{MEGII:2023ltw}
K.~Afanaciev \textit{et al.} [MEG II],
Eur. Phys. J. C \textbf{84} (2024) no.3, 216
[arXiv:2310.12614 [hep-ex]].

\bibitem{MEG:2016leq}
A.~M.~Baldini \textit{et al.} [MEG],
Eur. Phys. J. C \textbf{76} (2016) no.8, 434
[arXiv:1605.05081 [hep-ex]].

\bibitem{Belle:2021ysv}
A.~Abdesselam \textit{et al.} [Belle],
JHEP \textbf{10} (2021), 19
[arXiv:2103.12994 [hep-ex]].


\bibitem{Fernandez-Martinez:2016lgt}
E.~Fernandez-Martinez, J.~Hernandez-Garcia and J.~Lopez-Pavon,
JHEP \textbf{08} (2016), 033
[arXiv:1605.08774 [hep-ph]].

\bibitem{Agostinho:2017wfs}
N.~R.~Agostinho, G.~C.~Branco, P.~M.~F.~Pereira, M.~N.~Rebelo and J.~I.~Silva-Marcos,
Eur. Phys. J. C \textbf{78} (2018) no.11, 895
[arXiv:1711.06229 [hep-ph]].

\bibitem{Blennow:2023mqx}
M.~Blennow, E.~Fern\'andez-Mart\'\i{}nez, J.~Hern\'andez-Garc\'\i{}a, J.~L\'opez-Pav\'on, X.~Marcano and D.~Naredo-Tuero,
JHEP \textbf{08} (2023), 030
[arXiv:2306.01040 [hep-ph]].


\bibitem{MEGII:2018kmf}
A.~M.~Baldini \textit{et al.} [MEG II],
Eur. Phys. J. C \textbf{78} (2018) no.5, 380
[arXiv:1801.04688 [physics.ins-det]].

\bibitem{Belle-II:2018jsg}
E.~Kou \textit{et al.} [Belle-II],
PTEP \textbf{2019} (2019) no.12, 123C01
[erratum: PTEP \textbf{2020} (2020) no.2, 029201]
[arXiv:1808.10567 [hep-ex]].


%



\bibitem{Qin:2017aju}
Q.~Qin, Q.~Li, C.~D.~L\"u, F.~S.~Yu and S.~H.~Zhou,
Eur. Phys. J. C \textbf{78} (2018) no.10, 835
[arXiv:1711.07243 [hep-ph]].

\bibitem{Jueid:2023fgo}
A.~Jueid, J.~Kim, S.~Lee, J.~Song and D.~Wang,
Phys. Rev. D \textbf{108} (2023) no.5, 055024
[arXiv:2305.05386 [hep-ph]].

\bibitem{Barman:2022iwj}
R.~K.~Barman, P.~S.~B.~Dev and A.~Thapa,
Phys. Rev. D \textbf{107} (2023) no.7, 075018
[arXiv:2210.16287 [hep-ph]].

\bibitem{Aoki:2023wfb}
M.~Aoki, S.~Kanemura, M.~Takeuchi and L.~Zamakhsyari,
Phys. Rev. D \textbf{107} (2023) no.5, 055037
[arXiv:2302.08489 [hep-ph]].





\bibitem{Dam:2018rfz}
M.~Dam,
SciPost Phys. Proc. \textbf{1} (2019), 041
[arXiv:1811.09408 [hep-ex]].

\bibitem{FCC:2018byv}
A.~Abada \textit{et al.} [FCC],
Eur. Phys. J. C \textbf{79} (2019) no.6, 474

\bibitem{Han:2018znu}
X.~F.~Han, T.~Li, L.~Wang and Y.~Zhang,
Phys. Rev. D \textbf{99} (2019) no.9, 095034
[arXiv:1812.02449 [hep-ph]].

\bibitem{Barman:2021xeq}
R.~K.~Barman, R.~Dcruz and A.~Thapa,
JHEP \textbf{03} (2022), 183
[arXiv:2112.04523 [hep-ph]].

\bibitem{Athron:2021iuf}
P.~Athron, C.~Bal\'azs, D.~H.~J.~Jacob, W.~Kotlarski, D.~St\"ockinger and H.~St\"ockinger-Kim,
JHEP \textbf{09} (2021), 080
[arXiv:2104.03691 [hep-ph]].

\bibitem{Wang:2022yhm}
L.~Wang, J.~M.~Yang and Y.~Zhang,
Commun. Theor. Phys. \textbf{74} (2022) no.9, 097202
[arXiv:2203.07244 [hep-ph]].


\bibitem{tHooft:1972tcz}
G.~'t Hooft and M.~J.~G.~Veltman,
Nucl. Phys. B \textbf{44} (1972), 189-213


	
\bibitem{Alonso:2012ji}
R.~Alonso, M.~Dhen, M.~B.~Gavela and T.~Hambye,
JHEP \textbf{01} (2013), 118
[arXiv:1209.2679 [hep-ph]].
	
\bibitem{Barr:1990vd}
S.~M.~Barr and A.~Zee,
Phys. Rev. Lett. \textbf{65} (1990), 21-24
[erratum: Phys. Rev. Lett. \textbf{65} (1990), 2920]
	
\bibitem{Goudelis:2011un}
	A.~Goudelis, O.~Lebedev and J.~h.~Park,
	Phys. Lett. B \textbf{707} (2012), 369-374
	[arXiv:1111.1715 [hep-ph]].
	

\bibitem{Chang:1993kw}
D.~Chang, W.~S.~Hou and W.~Y.~Keung,
Phys. Rev. D \textbf{48} (1993), 217-224
[arXiv:hep-ph/9302267 [hep-ph]].
	
\bibitem{Davidson:2016utf}
S.~Davidson,
Eur. Phys. J. C \textbf{76} (2016) no.5, 258
[arXiv:1601.01949 [hep-ph]].

\bibitem{Borsanyi:2020mff}
S.~Borsanyi, Z.~Fodor, J.~N.~Guenther, C.~Hoelbling, S.~D.~Katz, L.~Lellouch, T.~Lippert, K.~Miura, L.~Parato and K.~K.~Szabo, \textit{et al.}
Nature \textbf{593} (2021) no.7857, 51-55
[arXiv:2002.12347 [hep-lat]].	


\end{thebibliography}
\end{document}